\documentclass[11pt]{article}
\usepackage[superscript]{cite}

\makeatletter
\renewcommand\@biblabel[1]{#1.}
\makeatother

\usepackage{times,amsmath,amsfonts}

\newenvironment{sciabstract}{%
\begin{quote} \bf}
{\end{quote}}

\usepackage[utf8]{inputenc}
\usepackage[english]{babel}
\usepackage{graphicx}% Include figure files
\graphicspath{{Fig_Main/}}
\usepackage{dcolumn}% Align table columns on decimal point
\usepackage{bm}% bold math
\usepackage[dvipsnames]{xcolor}
\newcommand{\modif}[1]{{\color{black}#1}}
\usepackage[normalem]{ulem}
\usepackage{siunitx}
\DeclareSIUnit\Molar{\textrm{M}}
\date{}

\begin{document}

\title{Intrinsic Luminescence Blinking from Plasmonic Nanojunctions}

\author{Wen Chen$^1$, Philippe Roelli$^{1,2}$, 
Aqeel Ahmed$^1$, Sachin Verlekar$^1$,\\ 
Huatian Hu$^3$, Karla Banjac$^4$, Magal\'{i} Lingenfelder$^4$,\\
Tobias J. Kippenberg$^2$, Giulia Tagliabue$^5$, Christophe Galland$^{1\ast}$\\
\\
% \email{chris.galland@epfl.ch}
$^{1}$\small{Ecole Polytechnique F\'{e}d\'{e}rale de Lausanne,}\\
\small{Laboratory of Quantum and Nano-Optics, Lausanne 1015, Switzerland}\\
\small{$^2$Ecole Polytechnique F\'{e}d\'{e}rale de Lausanne,}\\
\small{Laboratory of Photonics and Quantum Measurements, Lausanne 1015, Switzerland}\\
\small{$^3$The Institute for Advanced Studies, Wuhan University, Wuhan 430072, China}\\
\small{$^4$Ecole Polytechnique F\'{e}d\'{e}rale de Lausanne,}\\
\small{Max Planck-EPFL Laboratory for Molecular Nanoscience, Lausanne 1015, Switzerland}\\
\small{$^5$Ecole Polytechnique F\'{e}d\'{e}rale de Lausanne,}\\
\small{Laboratory of Nanoscience for Energy Technologies, Lausanne 1015, Switzerland}\\
\small{$^\ast$To whom correspondence should be addressed; E-mail: chris.galland@epfl.ch}
}

\baselineskip20pt

\maketitle 

\begin{sciabstract}

\section*{Abstract}
\modif{
Plasmonic nanojunctions, consisting of adjacent metal structures with nanometre gaps, can support localised plasmon resonances that boost light matter interactions and concentrate electromagnetic fields at the nanoscale. 
}
In this regime, the optical response of the system is governed by poorly understood dynamical phenomena at the frontier between the bulk, molecular and atomic scales.
\modif{
Here, we report ubiquitous spectral fluctuations in the intrinsic light emission from photo-excited gold nanojunctions, which we attribute to the light-induced formation of domain boundaries and quantum-confined emitters inside the noble metal.  
Our data suggest that photoexcited carriers and gold adatom – molecule interactions play key roles in triggering luminescence blinking. 
Surprisingly, this internal restructuring of the metal has no measurable impact on the Raman signal and scattering spectrum of the plasmonic cavity.
Our findings demonstrate that metal luminescence offers a valuable proxy to investigate atomic fluctuations in plasmonic cavities, complementary to other optical and electrical techniques.
}

\end{sciabstract}

\maketitle

\section*{Introduction}

Plasmonic nanojunctions formed by ultrathin dielectric spacers between two metals enable reaching the quantum limits of light confinement at visible and near-infrared frequencies, with a growing number of applications in molecular science, nanophotonics, quantum optics and nanoscale optoelectronics \cite{baumberg2019}. 
By inserting molecules or low-dimensional materials in plasmonic nanojunctions their intrinsic optical, electronic and vibrational properties can be investigated with unprecedented sensitivity \cite{xu1999,zhang2013a,zrimsek2017}. Furthermore, these properties can be modified by leveraging giant values of the Purcell-factor \cite{acuna2012,hoang2016a,parzefall2019,bogdanov2020}, optomechanical coupling rate \cite{roelli2016,benz2016a} or vacuum Rabi splitting \cite{santhosh2016} -- values that typically surpass those of dielectric cavities. 
The generation of photo-excited charge carriers inside the metal can be enhanced by the plasmonic resonance and field enhancement, with potential applications in photo-catalysis \cite{brongersma2015,narang2016,cortes2017} and nanoscale light sources \cite{parzefall2019}. 
Despite progress in developing plasmonic nanojunctions as a universal platform to engineer light-matter interaction at the nanoscale, the realisation of their full potential is hindered by a limited understanding of physical processes driven by the tightly confined optical fields at the atomic scale \cite{banik2012,benz2016a,zhang2013a,li2017a,zhang2018,shin2018,lindquist2019,carnegie2020}.
Moreover, the modification of plasmon damping \cite{foerster2019,lee2019} and charge carrier dynamics \cite{bauer2006,hartland2011} by metal-molecule interfaces and intrinsic grain boundaries \cite{assefa2020} can further complicate the understanding of plasmonic nanojunctions.

Illustrating the emerging opportunities in this field, the efficiency of intrinsic light emission from a noble metal under optical or electrical pumping can be enhanced by many orders of magnitudes thanks to the giant Purcell factor provided by plasmonic nanocavities \cite{hu2012,lumdee2014,huang2015b,lin2016,lin2017,cai2018}. 
This plasmon-enhanced metal photoluminescence (PL) enables an increasing number of applications in imaging and nano-science
 \cite{vandijk2005,zheng2012a,zhang2018c}. 
Although its underlying principles are still under debate \cite{hugall2015,haug2015,lin2016,mertens2017,cai2018},
it is generally accepted that both interband and intraband transitions in the noble metal contribute to the radiative recombination of photo-excited carriers, with their relative contributions determined by the bulk band structure \cite{boyd1986}, the electron-hole pair energy \cite{cai2018}, and the degree of spatial confinement \cite{brown2016}.
At the meso- to macroscopic scale ($\sim$ 10-100 nm) governing the plasmonic response, the band structure of the metal is bulk-like.
In contrast, at the atomic scale, studies of metal clusters and nanoparticles below a few nanometers have shown that quantum confinement leads to bright emission from discrete energy states, as well as from metal-ligand hybrid states \cite{zheng2004,shang2011,zheng2012a,li2017a}. 
To date, these two domains have been largely considered as separate realms.

\begin{figure*}[htp!]
    \centering
    \includegraphics[width=0.67\columnwidth]{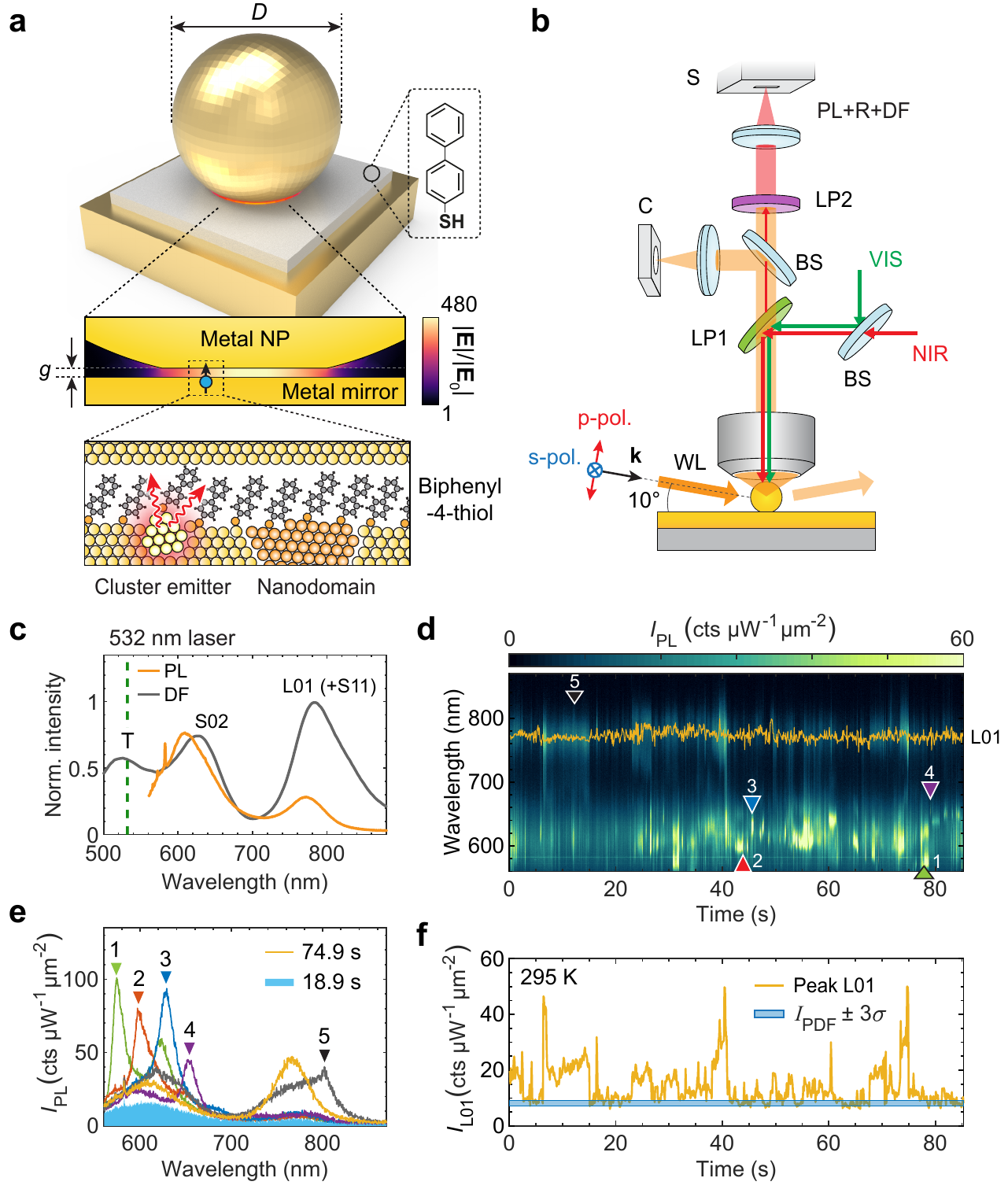}
    \caption{
    {\footnotesize 
    \textbf{Blinking of metal photoluminescence (PL) in a single nanojunction}.
   (\textbf{a}) Schematic representation of a nanojunction, made of a gold mirror, a self-assembled biphenyl-4-thiol (BPhT) monolayer (\textit{g} $\sim$1~nm) and a faceted gold nanoparticle (\textit{d} $\sim$80~nm). Middle inset: simulation of electric field distribution in the nanojunction region. Lower inset: illustration of the luminescent nano-clusters or nano-domains forming under laser irradiation, which we invoke as the cause of PL fluctuations shown in \textbf{d-f}.
    (\textbf{b}) Schematic of the optical setup enabling three simultaneous types of measurement: PL under 532~nm (VIS) excitation, Raman scattering (R) under tunable near-infrared (NIR) excitation, and dark-field scattering (DF) under p- or s-polarised grazing angle white light (WL) illumination. BS: beam-splitter, LP: long-pass filter, C: camera, S: spectrometer.
    (\textbf{c}) PL spectrum of a single nanojunction averaged over the entire duration of panel \textbf{d} (orange curve) and DF scattering spectrum of the same nanojunction (grey curve, calibrated by the illumination spectrum). Labeling of the modes is detailed in Fig.~\ref{fig:multimode}\textbf{g} and \textbf{h}.
    (\textbf{d}) Time series of  %normalized
    plasmon-enhanced PL (the color scale is saturated for better visibility of weak emission periods). 
    Power density of the 532~nm laser: \SI{\sim 70}{\micro\watt\per\square\micro\meter},
    %$\sim 70~\mathrm{\mu}\text{W} \mathrm{\mu}\text{m}^{-2}$, %$\sim 7\times 10^3~$W/cm$^2$, 
    camera exposure time = 0.1~s, numerical aperture: 0.85, room temperature. 
    (\textbf{e}) Individual examples of anomalous emission (referring to \textbf{d}) deviating from the typical baseline PL emission (shaded blue).
    (\textbf{f}) Time trace of the maximum PL intensity around the L01 mode, as marked by orange trace in \textbf{d}. The shaded blue area corresponds to the instrument-limited level of fluctuations, where $I_\mathrm{PDF}$ is the peak in the intensity probability density function (PDF) and $\sigma$ is the standard deviation of the measurement noise at this signal intensity (see Supplementary Fig.~7).
    }}
    \label{fig:PL}
\end{figure*}

In this article, we show that such a distinction should be reconsidered. 
We discover that the intrinsic light emission from gold plasmonic nanojunctions generally consists of two components:
(i) a stable light emission baseline, spectrally following the plasmonic resonances and governed by the bulk metal band structure, and 
(ii) a contribution from quantum-confined emitters and crystal defects randomly forming and disappearing near the metal surface (Fig.~\ref{fig:PL}a). %under the influence of molecules on the surface and photo-excited carriers based on the highly confined optical field 
This latter process, which results in a fluctuating (i.e. blinking) luminescence and is the focus of our study, has its origin at the atomic scale, but is made observable thanks to the Purcell effect provided by the plasmonic modes of the entire junction.
The Purcell-enhanced emission from quantum-confined metallic emitters transiently results in sharper linewidths (higher apparent Q-factors) and much higher quantum yields compared to the baseline emission.
Our findings reveal a phenomenology where luminescence blinking is due to metastable configurations of the atomic lattice, instead of fluctuations in the charge state as observed to date in molecular fluorophores and low-dimensional semiconductors \cite{bout1997,frantsuzov2008}.  They raise interrogations about the validity of using bulk electronic band structures to model chemical and photochemical interactions at the surface of plasmonic structures.%\sout{a new wave of} 
We anticipate that our results will motivate \modif{further}  experimental investigations of optically and electrically induced light emission from plasmonic nanojunctions, with specific attention devoted to metastable and transient states of emission and their relationship with modifications in the carrier relaxation pathways.

%%%%%%%%%%%%%%%%%%%%%%%%%%%%%%%%%%%%%%%%%%%%%%%%%%%%%%%%%%%%%%%%%%%%%%%%%%%%
%%%%%%%%%%%%%%%%%%%%%%%%%%%%%%%%%%%%%%%%%%%%%%%%%%%%%%%%%%%%%%%%%%%%%%%%%%%%

\section*{Results}

\subsection*{\modif{Blinking of metal PL in single plasmonic nanojunctions}}

We fabricated plasmonic nanojunctions following the ``nanoparticle-on-mirror" approach \cite{aravind1982,mubeen2012,chen2018,chen2018a}.
Starting from a metallic mirror (with thickness $>70$~nm), precise control of the spacer thickness was achieved by self-assembly of a molecular monolayer \cite{ulman2013introduction}, or by the transfer of a transition metal dichalcogenide (TMDC) monolayer \cite{chen2018}, or by the atomic layer deposition of an oxide \cite{chen2018a} -- or by a combination thereof.
Subsequently drop-casting nanoparticles of the desired shape and composition (diameter kept at $\sim 80$~nm in the following) resulted in the formation of nanojunctions with well-controlled metal spacing, and tailored optical resonances dominated by the excitation of localised surface plasmons with large field enhancement inside the gap (Fig.~\ref{fig:PL}a). 
In most cases, we encapsulated the final structures in a thin ($\sim 5$ to $10$~nm) aluminum oxide layer for improved long-term stability. 
Full details about sample fabrication and characterisation are presented in the Supplementary \modif{Methods}. 
Overall, we acquired hundreds of PL time traces on individual nanojunctions in more than 20 different samples with distinct mirror, spacer and nanoparticle compositions. 
In order to simultaneously collect vibrational Raman scattering and elastic Rayleigh scattering and to study the temperature dependence of blinking statistics, we built room-temperature and cryogenic multi-functional microscopes for single-particle spectroscopy, as schematically depicted in Fig.~\ref{fig:PL}b.
A complete list of fabricated samples and details of the setups are described in the Supplementary {Table 1 and Methods}.

\begin{figure*}[htp]
    \centering
   \includegraphics[width=1\columnwidth]{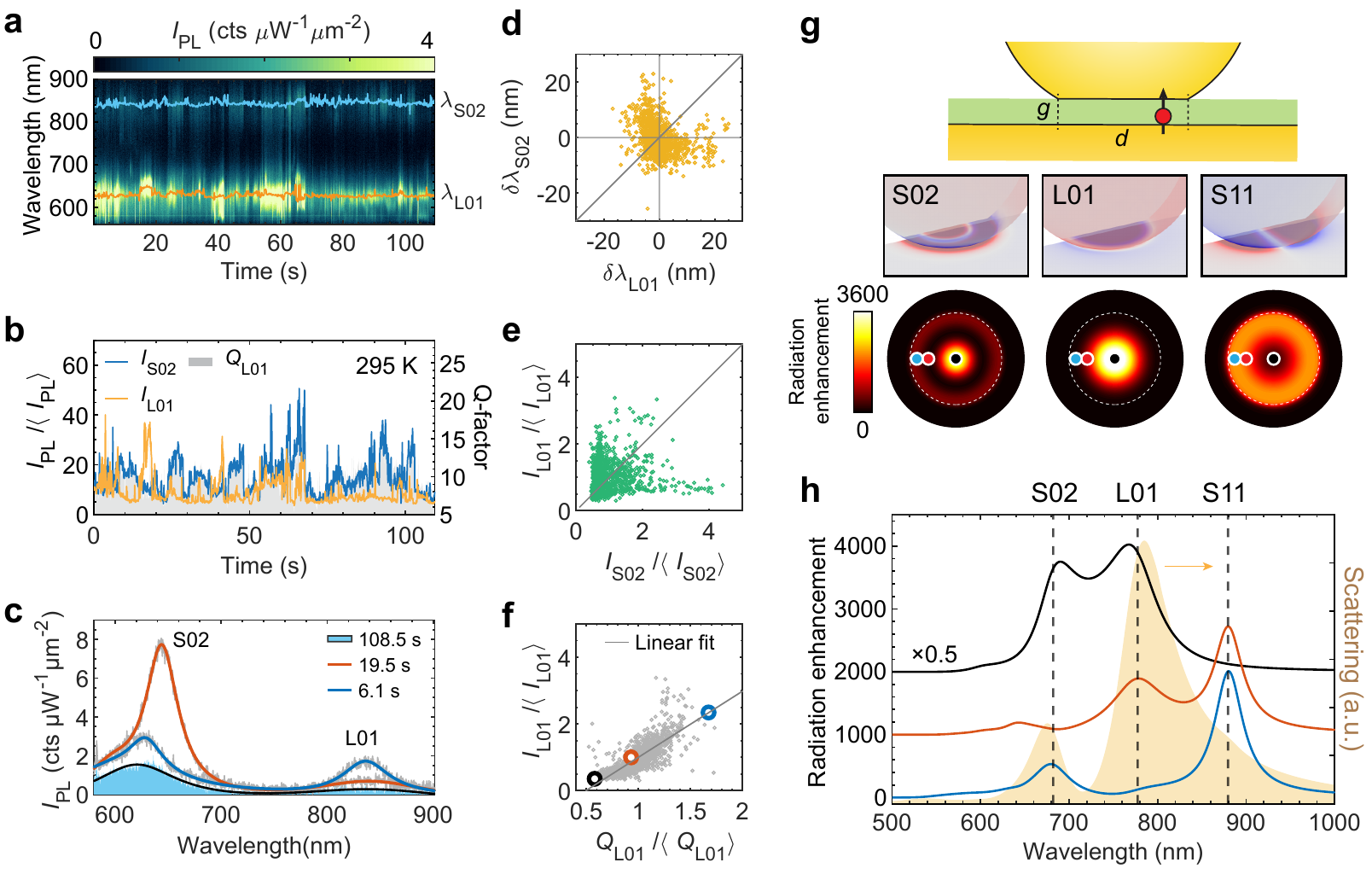}
    \caption{
    {\small
    \textbf{Multi-mode blinking: evidence for spatially localised fluctuating sources of emission. } 
    (\textbf{a}) Fluctuating PL time-trace from a nanojunction emitting from the L01 and S02 gap modes. Excitation power density at 532~nm: 
    \SI{\sim 45}{\micro\watt\per\square\micro\meter},
    %$\sim 45~\mu\text{W}/\mu\text{m}^2$
    , numerical aperture 0.85, exposure time: 0.1s, room temperature.
    (\textbf{b}) Time series of peak PL intensities at the L01 (blue curve) and S02 (orange curve) resonances along with the $Q$-factor of the L01 mode (grey-shaded area).
    (\textbf{c}) Examples of PL spectra with different $Q$-factors (as fitted by Lorentzian functions), along with the typical baseline PL spectrum (blue shaded area).
    (\textbf{d, e}) Distribution of the PL peak wavelength (\textbf{d}) and intensity (\textbf{e}) (relative to their time average denoted by brackets) around one resonance vs. the other, showing no correlations. 
    (\textbf{f}) Distribution of relative PL intensity vs. $Q$-factor for the L01 emission, showing a positive correlation. Individual spectra from \textbf{c} are highlighted with black, red and blue circles.
    (\textbf{g, h}) Full-wave simulation (finite-element modelling) of the optical response of a faceted nanojunction. 
   (\textbf{g}) Based on the surface charge distributions taken at resonance, the modes are identified as the lowest frequency Fabry–Pérot-like transverse cavity mode S11, the dipolar bonding antenna mode L01 and the higher-order cavity mode S02, respectively \cite{zhang2019}.
These modes feature distinct spatial distributions of their local photonic density of states (PDOS), which result in different far-field emission spectra (solid lines in \textbf{h}, offset for clarity) when a radiating point dipole is placed at the different locations shown by color-coded full circles in \textbf{g}. The yellow-shaded area in \textbf{h} corresponds to the calculated DF spectrum.
    }} 
    \label{fig:multimode}
\end{figure*}

We present first the results from a nanojunction consisting of a chemically synthesised gold flake with (111) surface, a self-assembled biphenyl-4-thiol (\modif{BPhT}) monolayer, and a commercially available colloidal gold nanoparticle (nominal size 80~nm) (Fig.~\ref{fig:PL}).
The plasmonic response of the single nanojunction is first characterised by dark-field (DF) scattering spectroscopy using white light excitation from the side at a glazing angle with tunable polarisation, so that specular reflection from the substrate is not collected by the objective lens (Fig.~\ref{fig:PL}b). 
Without specific mention, the DF measurements in the following are all using p-polarised white light.
The DF spectrum exhibits three major features (Fig.~\ref{fig:PL}c). The strong peak in the near-infrared is attributed to a longitudinal dipolar antenna mode (polarised normal to the substrate) with strong field enhancement in the gap (labeled L01 in Fig.~\ref{fig:PL}c). 
Additionally, when the diameter of the nanoparticle facet in contact with the spacer exceeds about $10$~nm, the structure supports Fabry–Pérot-like metal-insulator-metal gap modes. These may hybridize with the vertically polarised antenna modes \cite{tserkezis2015a,zhang2019} giving rise to higher-order modes labeled S02 (observed around $62$0~nm) and S11 (overlapping with L01 for this particular nanojunction). Finally, around $530$~nm, the transverse plasmon mode of the nanojunction (labeled T) can be observed. These attributions are confirmed by numerical calculations (see Fig.~\ref{fig:multimode}g,h) and polarisation-dependent DF measurements (Supplementary Fig.~9).

Efficient excitation of metal PL from the single nanojunction is achieved at a wavelength of 532~nm, when the photon energy matches optically-allowed interband ($d$ to $sp$ band) transitions in gold.
This energy is also resonant with the transverse mode, enhancing the absorption cross-section.
The PL emission from a nanojunction is much stronger than the weak continuum PL collected from the bare metal substrate (see Supplementary Fig.~8), despite the fact that the area of the nanojunction is at least 500 times smaller than our spot size. 
Here, the PL spectrum is the time-average of the series shown in Fig.~\ref{fig:PL}d. 
This demonstrates that PL from the metal is enhanced by orders-of-magnitude due to the combined effect of large near-field coupling to the nanocavity modes and efficient far-field coupling through the antenna effect \cite{bogdanov2020}.

When recorded with a short exposure time (0.1~s), the PL time-trace of the nanojunction features pronounced blinking and spectral wandering (Fig.~\ref{fig:PL}d, see Supplementary \modif{Movie} 1 for the entire time trace). Closer inspection of PL spectra at selected times (Fig.~\ref{fig:PL}e) reveals prominent intensity fluctuations of the L01 mode (orange curve) as well as the appearance of randomly-occurring bright PL emission lines around the S02 and S11 modes (green, red, blue, purple and gray curves). 
At all times, we also observe the presence of a persistent baseline emission, which corresponds to the weakest emission of the time series (blue shaded area in Fig.~\ref{fig:PL}e). 
This baseline PL is attributed to Purcell-enhanced radiative recombination of non-thermal excited carriers through both inter- and intra-band processes, as discussed in previous literature \cite{lumdee2014,huang2015b,cai2018}. 
The PL peak intensity around the L01 mode (Fig.~\ref{fig:PL}f, orange curve) exhibits prominent fluctuations lasting from milliseconds (see below) up to seconds, well beyond the $3\sigma$ interval of the calibrated measurement noise, which includes shot noise and technical noise (Fig.~\ref{fig:PL}f, blue-shaded area; see details in Supplementary Fig.~7).
PL blinking was also consistently observed at lower temperatures (see below, and Supplementary Fig.~16).

%%%%%%%%%%%%%%%%%%%%%%%%%%%%%%%%%%%%%%%%%%%%%%%%%%%%%%%%%%%%%%%%%%%%%%%%%%%%

%%%%%%%%%%%%%%%%%%%%%%%%%%%%%%%%%%%%%%%%%%%%%%%%%%%%%%%%%%%%%%%%%%%%%%%%%%%%

\subsection*{\modif{Evidence for light-induced fluctuating local emitters}}

To gain further insight into the origin of metal PL blinking, 
we analyze the spectral wandering and lineshape narrowing that accompany blinking, and study the correlations that may exist between fluctuations in emission wavelength, intensity and linewidth from different regions of the full spectrum. 
Fig.~\ref{fig:multimode}a displays another representative time trace with a typical multi-peak PL, with selected spectra shown in Fig.~\ref{fig:multimode}c.
For each mode, we track the wavelength of maximum PL (${\lambda}_\mathrm{L01}$ and ${\lambda}_\mathrm{S02}$), the peak intensity (${I}_\mathrm{L01~}$ and ${I}_\mathrm{S02}$) and the linewidth (expressed here in terms of $Q$-factor, with $Q_\mathrm{L01}$ shown as the shaded area in Fig.~\ref{fig:multimode}b). 
A first result of this analysis is that no significant (anti-)correlations exist between the peak wavelengths ${\lambda}_\mathrm{L01}$ and ${\lambda}_\mathrm{S02}$  (Fig.~\ref{fig:multimode}d), nor between the normalised peak intensities ${I}_\mathrm{L01}$ and ${I}_\mathrm{S02}$ (Fig.~\ref{fig:multimode}e).
Any model relying on the modification of the entire, mesoscopic plasmonic response would therefore be difficult to reconcile with our observations.

In contrast, we observe a clear positive correlation between the relative increases of emission intensity (${I}_\mathrm{L01}/\langle {I}_\mathrm{L01} \rangle$) vs. $Q$-factor (${Q}_\mathrm{L01}/\langle{Q}_\mathrm{L01}\rangle$, Fig.~\ref{fig:multimode}f). 
In other words, the higher the blinking PL intensity, the narrower the effective PL linewidth. 
From the point of view of traditional mechanisms proposed so far to describe plasmon-enhanced light emission from metal, such behavior is difficult to rationalize. 
Indeed, an increase in $Q$-factor together with increased radiation rate would reflect a reduction of the nonradiative plasmonic losses, and we are not aware of a mechanism that could lead to such drastic variations over millisecond time scales. 
All observations suggest instead that an atomic-scale mechanism is causing PL blinking, without affecting the overall plasmonic response.

To check if fluctuating point-like emitters could yield such a behavior, we implemented full-wave simulations of a nanojunction consisting of an 80 nm Au nanoparticle with facet diameter \textit{d} = 40 nm on a Au mirror with spacer thickness \textit{g} = 1.3 nm (Fig.~\ref{fig:multimode}g, see details in Supplementary \modif{Methods}). 
Under the same illumination and collection geometry as used in the DF measurement, the simulated scattering spectrum (shaded yellow curve in Fig.~\ref{fig:multimode}h) matches our experimental data (Fig.~\ref{fig:PL}c). 
Three localised gap plasmon modes S02, L01 and S11 can be identified from their distinctive surface charge distributions (middle 3 panels in Fig.~\ref{fig:multimode}g).
To emulate a randomly generated point-like emitter, we use a broadband, vertically oriented electric dipole placed on the metal surface at three different positions  (blue, red and black dots in 3 bottom panels of Fig.~\ref{fig:multimode}g). 
Different radiation enhancements, determined by the local photonic densities of states (PDOS) and radiation angular distribution, are thus probed depending on the overlap between the emitter position and the field distributions of the different gap modes (Fig.~\ref{fig:multimode}h).
From these simulations, we infer that spatially localised fluctuations in PL quantum yield are consistent with uncorrelated intensity fluctuations in different modes (Fig.~\ref{fig:multimode}e). 
However, this toy model assumes that the PL spectrum is governed by the local PDOS only; it fails to explain the magnitude of wavelength fluctuations (Fig.~\ref{fig:multimode}d) and changes in linewidth (Fig.~\ref{fig:multimode}f) that we observe in some instances, in particular for thiol-functionalised gold substrates. 
For other nanojunctions with purely inorganic spacers such alumina, PL blinking is less pronounced and is not accompanied by noticeable changes in peak wavelength and linewidth  (e.g. Supplementary Fig.~10b, 10c and 10f). 
In the following, we focus our attention on the more pronounced blinking features characteristic of organic spacers.

To accommodate our observations, we propose that bright emission centers, consisting of nanoscale metallic domains and/or metal atom clusters, are being formed in the metal surface layer during laser irradiation.
Their optical transitions are dominated by quantum-confined electronic states within the $s-p$ band of gold \cite{zheng2004,shang2011,zheng2012a} (possibly hybridized with electronic states of the spacer material, in particular through their sulfur atoms).
This model is fully consistent with the results shown in Fig.~\ref{fig:multimode}a-f: 
isolated gold clusters or very small nanoparticles \cite{zheng2004,shang2011,zheng2012a,li2017a}
are capable of generating PL emission with a wide range of quantum yields, lifetimes and center wavelengths (covering visible and near-infrared), determined by their size and metal-ligand interaction.
In our structures, the plasmonic modes provide a large Purcell-enhancement (Fig.~\ref{fig:multimode}g,f) which makes the blinking emission predominant close to the plasmonic resonances observed in DF and in the baseline PL. 
Moreover, if we attribute the brightest emission periods to quantum-confined states in nano-clusters, their linewidths are expected to be narrower than that of the plasmon, as we do observe (Fig.~\ref{fig:multimode}c,f). 
Their emission wavelengths could also be affected by local charging \cite{borys2009} through the DC Stark effect.

A related mechanism is the temporary formation of new grain boundaries and other localised lattice defects, which can scatter electrons and are thus expected to relax wave-vector conservation, leading to a local increase in intraband radiative recombination rate \cite{brown2016} -- but without a reduction in emission linewidth nor a shift in emission wavelength. 
This mechanism explains well the moderate intensity blinking we observe in all constructed nanojunctions, irrespective of their surface chemistry and spacer material (see Supplementary Fig.~10 and 11). 
For completeness, we mention that the formation of charge-transfer states \cite{lee2019} or surface dipoles \cite{foerster2019} has been shown to increase the electron scattering rate, possibly providing an alternative or complementary explanation for PL blinking.

%%%%%%%%%%%%%%%%%%%%%%%%%%%%%%%%%%%%%%%%%%%%%%%%%%%%%%%%%%%%%%%%%%%%%%%%%%%%
\begin{figure}[htp]
    \centering
   \includegraphics[width=0.6\columnwidth]{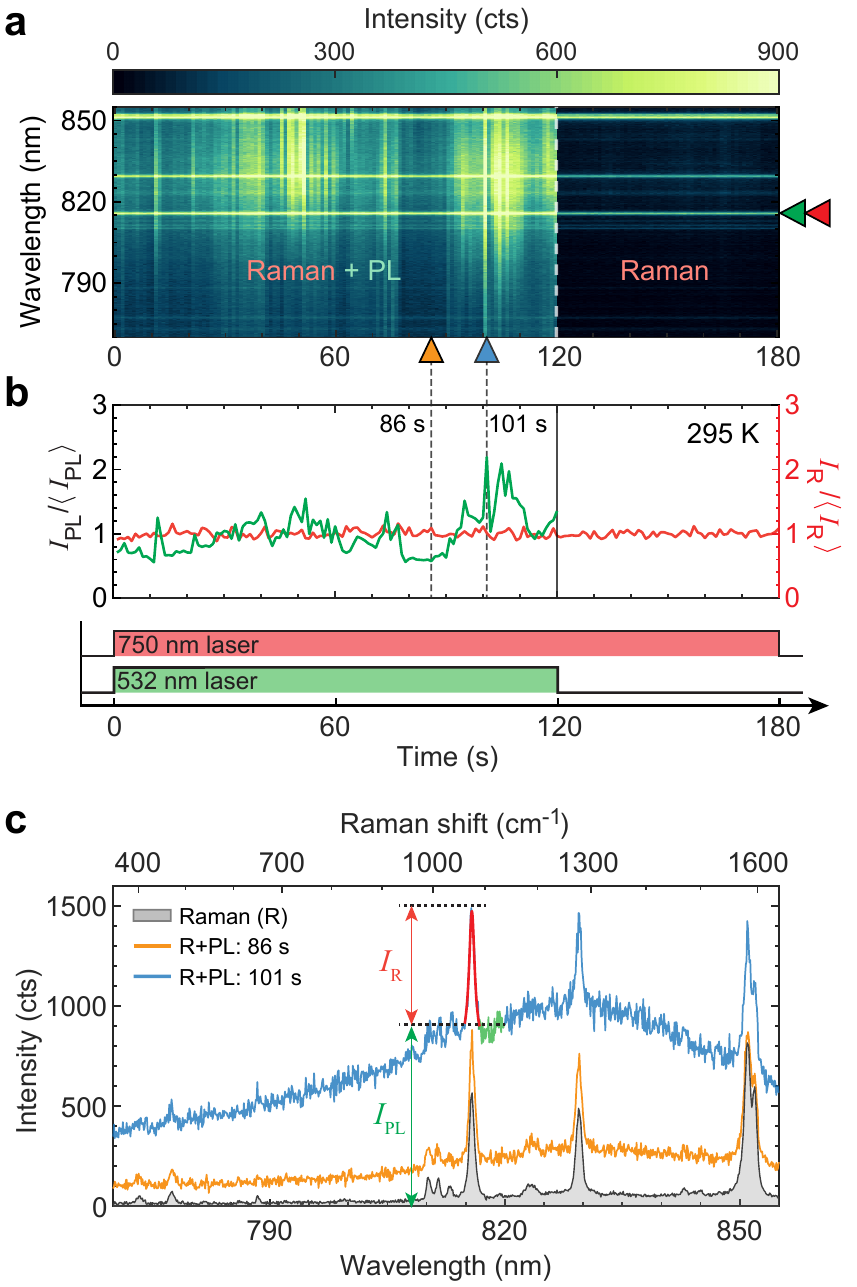}
    \caption{
    {\small 
    \textbf{Blinking PL with stable plasmon-enhanced Raman spectra.} 
	(\textbf{a}) Time series of emission spectra acquired under dual excitation with 532~nm and 750~nm laser beams (first 120~s) with respective power densities \SI{\sim 200}{\micro\watt\per\square\micro\meter},
	%$\sim 200~\mu\text{W}/\mu\text{m}^2$ 
	and \SI{\sim 10}{\micro\watt\per\square\micro\meter},
	%$\sim 10~\mu\text{W}/\mu\text{m}^2$
	, and then with 750~nm excitation alone (after 120~s). Camera exposure time = 1~s; numerical aperture: 0.95; room temperature. {The} color scale is saturated for better visibility of the fluctuations. 
    (\textbf{b}) Time series of the PL intensity (cf. $\textit{I}_{\mathrm{PL}}$ in \textbf{c}) and PL-subtracted Raman intensity (cf. $\textit{I}_{\mathrm{R}}$ in \textbf{c}), normalised by their respective time-averages ($\langle\textit{I}_{ \mathrm{PL}} \rangle$ and $\langle \textit{I}_{\mathrm{R}}\rangle$).
    (\textbf{c}) Example of individual Raman+PL spectra (blue and orange curves) and time-averaged Raman spectrum under 750~nm excitation alone (120 to 180~s, gray area). 
    The green and red-substituted points on the blue curve highlight how $\textit{I}_{\mathrm{PL}}$ and $\textit{I}_{\mathrm{R}}$ are defined.
    }}
    \label{fig:PL_Raman}
\end{figure}
%%%%%%%%%%%%%%%%%%%%%%%%%%%%%%%%%%%%%%%%%%%%%%%%%%%%%%%%%%%%%%%%%%%%%%%%%%%%

\subsection*{\modif{PL Blinking with stable Raman spectrum}}

In contrast to predictions from existing models proposed to explain fluctuations in surface-enhanced Raman scattering  \cite{banik2012,benz2016a,sprague-klein2017,shin2018,lindquist2019} and background emission in plasmonic nanojunctions \cite{carnegie2020} (a more comprehensive literature review is provided in the Supplementary \modif{Note 1}), we cannot relate the PL blinking to fluctuations of field enhancement inside the gap, nor to changes in the plasmonic response, as we now demonstrate. 
To obtain an independent probe of the local field enhancement, while simultaneously monitoring PL blinking, we performed two-tone excitation with both a 532~nm laser to efficiently generate PL, and with another continuous-wave laser tuned at 750~nm so that the Stokes vibrational Raman signal from the \modif{BPhT} molecules embedded in the gap is resonant with a near-infrared plasmonic mode (Fig.~\ref{fig:PL_Raman}). 
If blinking were caused by fluctuations in local field enhancement, such fluctuations would be reflected, at least in part, on the Raman signal \cite{banik2012,benz2016a,shin2018,chen2018}, since molecules are thought to occupy the entire gap region.
As a representative example, Fig.~\ref{fig:PL_Raman}a shows time series of the Raman+PL (first 120~s) and sole Raman spectra from a nanojunction, with selected Raman+PL and time-averaged Raman (last 60~s) spectra shown in Fig.~\ref{fig:PL_Raman}c.
Remarkably, the fluctuations of the Raman signal ($I_\mathrm{R}$, Fig.~\ref{fig:PL_Raman}c) remain within the irreducible measurement noise, while much more pronounced fluctuations of the underlying PL emission ($I_\mathrm{PL}$, Fig.~\ref{fig:PL_Raman}c) are observed. 
This measurement (which was repeated on many nanojunctions with the same result) provides evidence that the near-field enhancement and thereby the local density of photonic states remain stable during PL blinking -- in stark contrast with previous observations of fluctuating Raman scattering, e.g. \cite{xu1999,lindquist2019}.
Moreover, based on this observation, we conclude that mechanisms which can be sensitively probed by Raman scattering, including chemisorption \cite{banik2012}, adsorbate-metal charge transfer and charging effects \cite{sprague-klein2017}, are unlikely to be the dominant cause of PL blinking (see detailed discussion in Supplementary \modif{Note 1}). 
Last, if there were a build-up of high DC fields across the gap it should result in a DC Stark shift of the Raman peaks, which we do not observe.

We note that under 750~nm excitation alone, the absence of interband transitions in gold strongly reduces the PL excitation cross-section, and we typically observe a very low amount of PL -- except for the brightest blinking events (Supplementary Fig.~13), akin to the so-called `flares' reported in \cite{carnegie2020}.
We also occasionally observe the appearance of many new Raman sidebands, with intensities more than ten times above the normal Raman signal (see Supplementary Fig.~14).
Recent reports have invoked the formation of `picocavities' \cite{benz2016a,shin2018} to explain such events, which are proposed to be related to metal protuberances causing atomic scale confinement of light. 
Our measurements show that PL blinks independently of such unusual Raman events, confirming that a new mechanism is at play during PL blinking.

%%%%%%%%%%%%%%%%%%%%%%%%%%%%%%%%%%%%%%%%%%%%%%%%%%%%%%%%%%%%%%%%%%%%%%%%%%%%
\begin{figure}[htp]
    \centering
   \includegraphics[width=0.7\columnwidth]{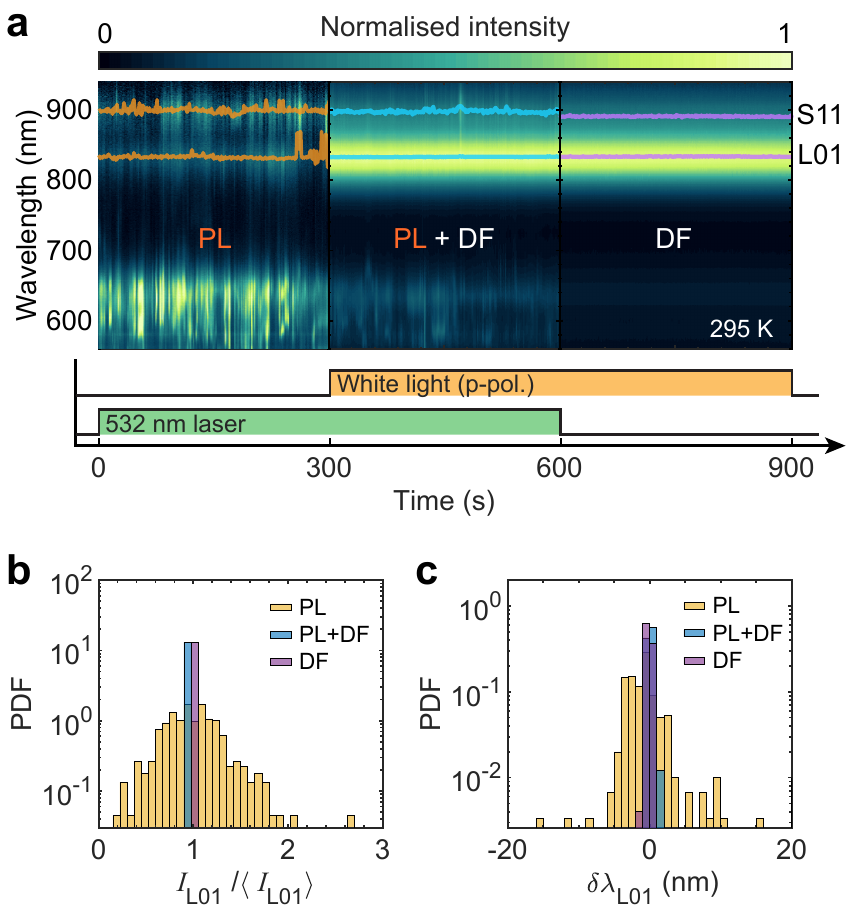}
    \caption{
    {\small 
    \textbf{Blinking PL with stable dark-field.} 
    (\textbf{a}) Spectral time series from an individual nanojunction under sequential illuminations with 532~nm laser alone (PL), both 532~nm laser and white light (PL+DF, normalised by illumination spectrum) and white light alone (DF, normalised by illumination spectrum). The positions of maximum emission close to the L01 and S11 modes are shown as orange, blue and purple traces for the PL, PL+DF and DF regions, respectively.
    (\textbf{b, c}) Probability density functions (PDFs) for the relative L01 peak intensity (\textbf{b}) and wavelength shift (\textbf{c}) extracted from the PL (orange), DF (purple) and DF+PL (blue) regions in \textbf{a}. The DF+PL (blue) area in \textbf{b} are slightly offset away from the value of 1 to show it clearly.
     Experimental parameters: objective numerical aperture = 0.8; laser power density \SI{\sim 45}{\micro\watt\per\square\micro\meter},
    %$\sim 45~\mu\text{W}/\mu\text{m}^2$
    ; white light is $p$-polarised; exposure time = 1s, room temperature. 
    }}
    \label{fig:PL_DF}
\end{figure}
%%%%%%%%%%%%%%%%%%%%%%%%%%%%%%%%%%%%%%%%%%%%%%%%%%%%%%%%%%%%%%%%%%%%%%%%%%%%

\subsection*{\modif{PL Blinking with stable dark-field scattering spectrum}}

Next, we design an experiment to verify that the DF scattering spectrum, which sensitively depends on nanoparticle shape and gap size \cite{tserkezis2015a,baumberg2019}, remains stable over time under green light excitation while PL blinks (Fig.~\ref{fig:PL_DF} and Supplementary \modif{Movie} 2). 
Figure ~\ref{fig:PL_DF}a shows spectral time series from a single nanojunction under sequential illuminations with 532~nm laser alone; together with white light; and with white light alone.
To allow quantitative comparison between the DF and PL fluctuations we plot the probability density functions (PDFs) of the L01-related peak intensity (Fig.~\ref{fig:PL_DF}b) and peak wavelength (Fig.~\ref{fig:PL_DF}c).
While the PL features strongly fluctuating intensity and peak wavelength, the elastic scattering of white light is highly stable, even when the laser is simultaneously exciting the nanojunction (PL+DF in Fig.~\ref{fig:PL_DF}a). 
Therefore, we conclude that rapid changes in nanoparticle shape cannot be the cause of PL blinking. 
Similarly, the model proposed in \cite{carnegie2020}, which invokes defects in the metal that alter the plasmonic resonance, also fails to agree with our measurements, since it predicts fluctuations of the elastic scattering spectrum correlated with brighter emission. 
Additionally, the stable plasmonic response in our system also excludes the appearance of quantum-tunnelling-induced charge-transfer plasmons, electro-luminescence of which was used to explain the broadband fluctuations in Ref. \cite{banik2012}.

\begin{figure*}[htp]
    \centering
   \includegraphics[width=0.7\columnwidth]{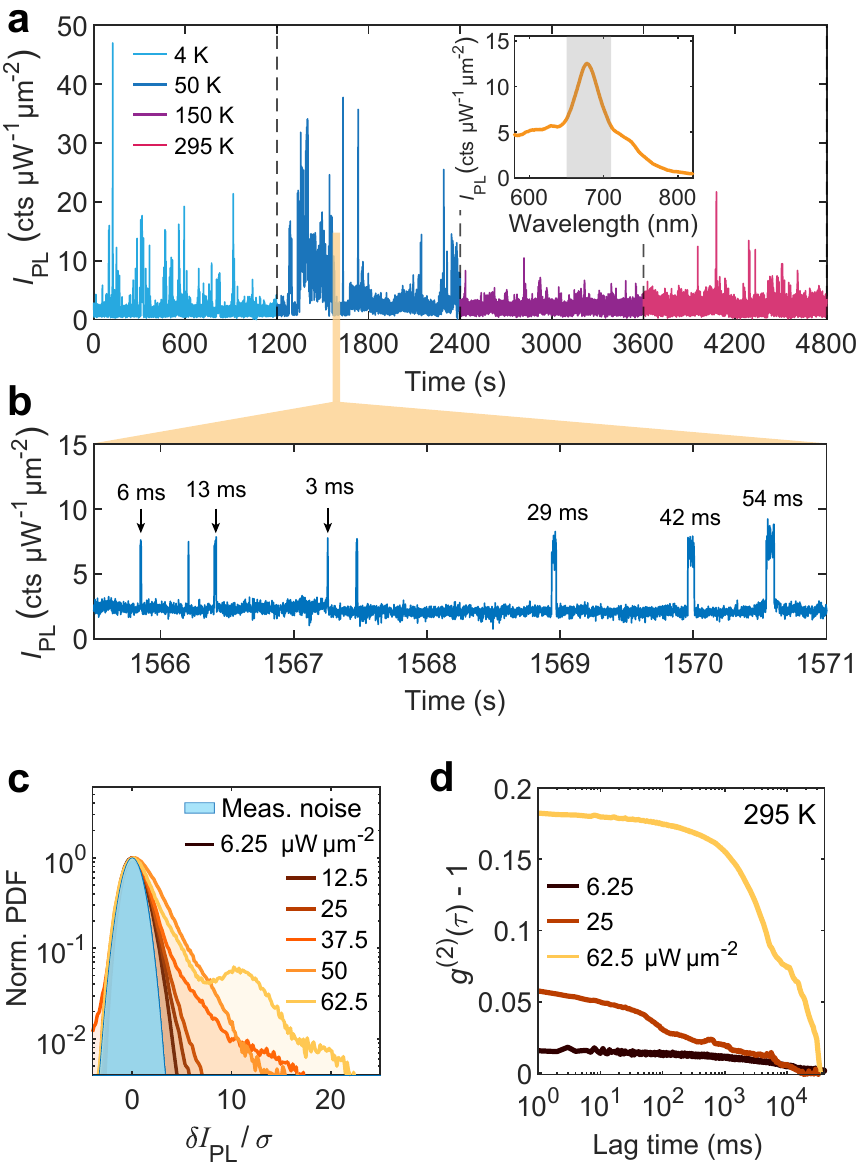}
    \caption{
    {\small
    \textbf{PL blinking as a function of laser power and sample temperature. } 
    (\textbf{a}) Temperature-dependent time series of PL intensity from a nanojunction, with the time-averaged PL spectrum shown in the inset. The PL signal was measured by a single photon counting module after spectral filtering (range shown in inset). Binning time, 1~ms. Excitation power density: \SI{\sim 50}{\micro\watt\per\square\micro\meter},
    %$\sim 50~\mu\text{W}/\mu\text{m}^2$.  
    (\textbf{b}) Enlarged view of \textbf{a} revealing shorter blinking events. More data presented in Supplementary Fig.~16 suggests that no clear relationship exists between temperature and blinking statistics.  
    (\textbf{c}) Probability density functions (PDFs) of peak PL intensity as a function of excitation intensity (room temperature), plotted against the re-scaled intensity $\delta I_\mathrm{PL}/\sigma$ to enable comparison. Here $\delta I_\mathrm{PL}$ represents the PL intensity deviation from the peak of the PDF; $\sigma$ is the standard deviation of the measurement noise at this signal level. 
    The PDFs are all centered around the averaged PL of the `OFF' state and normalised to the respective measurement noise (blue area). The corresponding time series can be found in Supplementary Fig.~17. 
    (\textbf{d}) Autocorrelation function of the PL intensity trace at different excitation intensities, evidencing increased fluctuations at higher laser powers.
    }} 
    \label{fig:temperature_power}
\end{figure*}

%%%%%%%%%%%%%%%%%%%%%%%%%%%%%%%%%%%%%%%%%%%%%%%%%%%%%%%%%%%%%%%%%%%%%%%%%%%%

\subsection*{\modif{Dependence of PL blinking on temperature and laser power}}

Finally, we turn our attention to the possible mechanisms for the formation and disappearance of the localised emitters responsible for blinking. 
We measured PL as a function of sample temperature and laser power (Fig.~\ref{fig:temperature_power}). 
To obtain a larger temporal dynamic range, we employed a single photon counting module (behind suitable filters to select emission from one plasmonic mode, see the inset of Fig.~\ref{fig:temperature_power}a). 
Figure ~\ref{fig:temperature_power}a shows the time series of the emission intensity from a single nanojunction at a sample temperature varying from 4~K up to room temperature, where the measured counts are summed into 1~ms time bins. 
From the enlarged view (Fig.~\ref{fig:temperature_power}b) we clearly identify the stable baseline PL intensity together with much brighter events, many of them lasting for few milliseconds only.
Even though Fig.~\ref{fig:temperature_power}a displays more frequent bright events at low temperature, we could not confirm any general relationship between the sample temperature and the blinking statistics in the range of 4~K to 300~K, as illustrated in Supplementary Fig.~16 by measurements performed on a larger number of nanojunctions.
Consequently, we can reject the hypothesis that the generation of localised emitters is thermally activated -- even though longer-lasting bright events seem more likely to be observed at lower temperature, suggesting that the relaxation to the baseline state may have a thermal component.
Multi-physics simulations (see Supplementary Fig.~19) confirm this conclusion by showing a rise in temperature due to laser illumination of a few Kelvin only -- negligible compared to the variation of bath temperature explored in Fig.~\ref{fig:temperature_power}a and b.

In Fig.~\ref{fig:temperature_power}c and d, we present the emission statistics as a function of excitation power for a fixed sample temperature (295 K). 
We find that blinking is hardly observable at the lowest excitation intensities (below \SI{\sim 10}{\micro\watt\per\square\micro\meter},
%$\sim 10~\mu\text{W}/\mu\text{m}^2$)
, where the stable baseline emission intensity is well above the dark count level with fluctuations barely exceeding the irreducible measurement noise (blue area in Fig.~\ref{fig:temperature_power}c). 
In contrast, as the laser intensity is increased, PL blinking is activated and becomes more pronounced and frequent, as illustrated by the power-dependent probability density functions (PDFs) plotted in Fig.~\ref{fig:temperature_power}c. 
These observations are confirmed by computing the autocorrelation (Fig.~\ref{fig:temperature_power}d) of the PL intensity traces under different laser powers, evidencing a higher level of PL fluctuation at higher laser powers.
This result highlights the key role of local optical field strength in activating the localised blinking emitters.

\section*{\modif{Discussion}}

We also find that blinking is much more likely to be activated under 532~nm excitation than at a longer wavelength beyond 600~nm (cf. Supplementary Fig.~13). 
It is even possible to activate blinking with temporary 532~nm illumination, and see the persisting increased level of background luminescence probed by a near-infrared laser immediately afterwards (see Supplementary Fig.~15). 
While we cannot totally exclude the role of near-field optical forces acting within the nanojunction \cite{juan2011,zhang2018},
this observation points to the key contribution of photo-excited electron-hole pairs in inducing the lattice restructuring, since 532~nm is close to the onset of interband absorption in gold.

Previous studies of metals under pulsed laser excitation have demonstrated the existence of a ``blast force" due to non-equilibrium electrons, which may deform the metal lattice \cite{chen2002}. 
On the other hand, it was demonstrated that hot electron injection upon a voltage pulse across a tip-surface junction can result in the restructuring of Au(111) surface in the presence of molecules at cryogenic temperatures \cite{merino2020}, which establish a possible link between the existence of hot carriers and the formation of atomic surface defects.
More theoretical work should be performed to determine whether similar forces can be relevant under weak continuous-wave excitation of nanoscale plasmonic cavities.
Based on simulations (Supplementary Fig.~20) we estimate that on the order of one photon per picosecond is absorbed by the nanojunction under typical excitation powers used here.%\modif{\sout{extremely}}
It is also possible that adatom-molecule complexes may cause  localised relaxation centers and hence enhance the probability for a local transfer of energy between non-thermal carriers and the lattice -- via a mechanism that could share similarities with electromigration induced by DC currents \cite{park1999}.
In this way, an energy as large as 2.4~eV per photon may be transferred to the lattice on the atomic scale.

Before concluding, we emphasize that blinking appears to be general; it could be observed consistently in many different samples. 
We investigated the impact of nanojunction composition on blinking by fabricating and characterising more than 20 different types of nanojunctions, as summarized in Table S1.
We systematically changed the substrate type, the spacer layer, and the nanoparticle material and shape, while maintaining similar plasmonic resonance frequencies and mode volumes.
A general conclusion can be drawn from these measurements (see Supplementary Figs. 10 and 11): while the molecules alone are not the source of photoluminescence, the magnitude and prevalence of PL blinking are indeed influenced by the spacer material and metal surface chemistry (in particular on the substrate side), with molecular spacer yielding more prominent blinking.
It could happen for at least two reasons: First, the stability and mobility of surface metal atoms depend on their direct environment, with molecular groups such as thiols perturbing the atomic arrangement in their vicinity and possibly facilitating light-induced restructuring. 
Second, molecules surrounding the metal can alter its local electron density via charge transfer \cite{foerster2019,lee2019}, which could favor electron-lattice scattering near the surface.
Finally, we characterise the optical response of single Au nanoparticles on SiO$_2$ (Supplementary Fig.~12). In this case, a weak PL blinking event was observed, but the occurrence is even rarer than that found in almost all the nanojunctions. It demonstrates that the existence of a gap mode with a strong local field is essential to the emergence of PL blinking and its observation.

\vspace{2cm}

In conclusion, we investigated the intrinsic photoluminescence (PL) blinking from plasmonic nanojunctions with various compositions, and obtained new insights into the origin of this phenomenon.
PL blinking is activated by the excitation laser and persists from room temperature down to 4~K. 
Bright PL events last from milliseconds up to minutes at low temperature. They can feature linewidths sharper than the plasmon resonance, and wandering peak wavelengths. 
This behavior contrasts with the nanojunction's baseline emission, its plasmonic response and local field enhancement, all of which remain stable while PL blinks.
These observations can be well explained by the proposed model: metastable localised quantum-confined emitters are photo-induced near the metal surface, and the fluctuating emission is enhanced by near-field coupling to the plasmonic antenna modes. 
The energy responsible for this lattice restructuring is not of thermal or ohmic origin; instead, we think it is deposited during the ultrafast relaxation of non-thermal photo-excited carriers. 
This rich physics occurs under weak continuous-wave excitation at tens of microwatts incoming power only, demonstrating the dramatic effect of plasmonic confinement on carrier and lattice dynamics in nanojunctions. 
Our results \modif{promote the} studies \modif{ of} plasmonic nanojunctions \modif{in a poorly understood regime}, involving phenomena at the interface between the atomic and mesoscopic scale.
Moreover, our findings raise questions regarding the microscopic mechanisms governing light emission from plasmonic nanojunctions, impacting their applications as nanoscale emitters. 
Finally, our work demonstrates that gap plasmons form the basis for new classes of materials whose optoelectronic properties are strongly modified by atomic-scale phenomena driven by non-thermal carriers.

\section*{Acknowledgments}

This works was funded by the Swiss National Science Foundation (SNSF) (project number PP00P2-170684), 
the European Research Council's (ERC) Horizon 2020 research and innovation programme under QTONE (grant agreement No. 820196), and the European Union H2020 research and innovation programme under THOR (grant agreement No. 829067).
The authors acknowledge Hongxing Xu for valuable comments and the IPHYS Characterisation Platform at EPFL for assistance.
P.R. acknowledges support from the Max Planck-EPFL Center for Molecular Nanoscience and Technology and from the European Research Council (ERC) under the European Union H2020 research and innovation programme (grant agreement No. 732894).

\section*{Author contributions}

W.C. and C.G. conceived the study; W.C., P.R., A.A. and S.V. performed the experiments; W.C., P.R. and S.V. analysed the data; H.H. performed the simulations; W.C., P.R., G.T. and C.G. wrote the manuscript. T.J.K. contributed to early ideas that led to this study. M.L. and K.B. performed high-resolution surface studies and proposed some mechanisms for local surface reconstruction.

\section*{Competing interests}
The authors declare no competing interests.

\section*{Data Availability}
The data that support the findings of this study are available in Zenodo repository with the digital object identifier (DOI): 10.5281/zenodo.4533192 ({https://zenodo.org/record/4533192}).

\section*{Code Availability}
The codes used to analyse and plot the data presented in this study are available from the authors upon reasonable request. 

%\bibliography{PL_Blinking_Final}% Produces the bibliography via BibTeX.
%\bibliographystyle{naturemag}

%%%%%%%%%%%%%%%%%%%%%%%%%%%%%%%%%%%%%%%%%%%%%%%%%%%%%%%%%%%%%%%%%%%%%%%%%%%%%%%%%%%%%%%%%%%%%%%%%%%%%%%%%%%%%%%%%%%%%%%%%%%%%%%%%%%%%%%%%%%%%%%%%%%%%%%%%%%%%%%%%%%%%%%%%%%%%%%%%%%%%%%%%%%%%%%%%%%%%%%%%%%%%%%%%%%%%%%%%%%%%%%%%%%%%%%%%%%%%%%%%%%%%%%%%%%%%%%%%%%%%%%%%%%%%%%%%%%%%%%%%%%%%%%%%%%%%%%%%%%%%%%%%%%%%%%%%%%%%%%%%%%%%%%%%%%%%%%%%%%%%%%%%%%%%%%%%%%%%%%%%%%%%%%%%%%%%%%%%%%%%%%%%%%%%%%%%%%%%%%%%%%%%%%%%%%%%%%%%%%%%%%%%%%%%%%%%%%%%%%%%%%%%%%%%%%%%%%%%%%%%%%%%%%%%%%%%%%%%%%%%%%%%%%%%%%%%%%%%%%%%%%%%%%%%%%%%%%%%%%%%%%%%%%%%%%%%%%%%%%%%%%%%%%%%%%%%%%%%%%%%%%%%%%%%%%%%%%%%%%%%%%%%%%%%%%%%%%%%%%%%%%%%%%%%%%%%%%%%%%%%%%%%%%%%%%%%%%%%%%%%%%%%%%%%%%%%%%%%%%%%%%%%%%%%%%%%%%%%%%%%%%%%%%%%%%%%%%%%%%%

\newpage
%\renewcommand\thefigure{\thesection\arabic{figure}}    
%\section*{SUPPLEMENTARY MATERIAL}
\setcounter{figure}{0}  
\renewcommand{\figurename}{\textbf{Supplementary Figure}}
\renewcommand{\tablename}{\textbf{Supplementary Table}}

\section*{Supplementary Methods}

\rule{0pt}{4ex}

\subsection*{Sample fabrication and characterisation}

Twenty-one types of plasmonic nanojunctions (see the list in Supplementary Table~\ref{tab:Sample_list}) were fabricated by a 4-step process: (1) fabrication of a metal film, (2) deposition of one or multiple layers on the metal film as a spacer, (3) drop-casting of metal nanoparticles on top of the spacer, and (4) growth of a compact Al$_2$O$_3$ layer on the sample surface for protection, the details of which are given in the following subsections. 

\subsubsection*{Metal film fabrication} 
Four types of metal films were prepared: evaporated Au and Cr (AuCr) film, template-stripped gold (TSAu) and silver (TSAg) films, colloidal Au microplates (AuMPs). The AuCr film with root-mean-square (RMS) roughness of $\sim$1.6 nm \cite{chen2018a} was fabricated by evaporating 5-nm-thick Cr film on a Si wafer, followed by 100-nm-thick Au layer with growth rate of 0.5 nm $\mathrm{s}^{-1}$. To fabricate the TSAu film, 200-nm-thick gold film was firstly evaporated on a clean Si wafer using electron beam evaporation at the same deposition rate. Next, the Au surface was glued with pieces of $\sim$1 cm$^2$ glass using an optical adhesive (NOA61), cured by ultraviolet light. Then the Si and glass slices were peeled off by a razor, leaving fresh ultrasmooth Au surface on the glass substrate. The TSAg films were fabricated in the same way. 
Atomic force microscopy was used to characterise the surface roughness and crystal grain size distribution of template-stripped (TSAu) and as-evaporated AuCr films, shown in Supplementary Fig.~\ref{fig:AFM}.
AFM images were collected by Dimension Fast Scan AFM (Bruker) coupled with Nanoscope V Controller (Bruker). FastScanB probes (Bruker) were used. All images were recorded in the air. Grain size analysis was conducted using Gwyddion software.

AuMPs were synthesized according to the method reported in Ref 2.
%\cite{kan2006}. 
Briefly, 6 mL ethylene glycol (Sigma-Aldrich) was firstly added into a 100 mL flask under 150 $^{\circ}$C oil bath. Then 1~mL of 0.2~M gold(III) chloride hydrate (HAuCl$_4$, Sigma-Aldrich) aqueous solution was injected into the flask. Next, 3~mL ethylene glycol solution containing 0.666~g of dissolved polyvinylpyrrolidone (PVP, Mw = 40000, Sigma-Aldrich) was dropped into the flask. After 10~min, AuMPs start to appear in the solution. The reaction was terminated after another 20 min, and the AuMPs were cleaned by acetone and ethanol solution, and finally stored in ethanol solution. The average lateral size and thickness are \SI{\sim 40}{\micro\meter} and 100~nm, respectively. 

\subsubsection*{Spacer fabrication} 

Different nanomaterials, including self-assembled monolayer (SAM) biphenyl-4-thiol (BPhT), monolayer  MoS$_2$, Al$_2$O$_3$, as well as native ligands on the colloid crystals, and their combinations, were used as spacers for the plasmonic nanojunctions, with thicknesses varying from ~0.7 nm to 3 nm (Supplementary Table~\ref{tab:Sample_list}).
SAM of BPhT molecules on the metal films with approximately 1~nm average thickness were obtained by immersing a fresh metal film in a BPhT ethanol solution with different incubation conditions. For AuCr, TSAu and TSAg films, 1~mM BPhT ethanol solution was used at room temperature for 2 hours incubation. In the case of the AuMPs, their solution was firstly drop-casted on a clean Si wafer and dried with nitrogen gas. Then the AuMP sample was immersed in a high concentration (0.1~M) BPhT solution at 70$^{\circ}$C for 24 hours. 

After the incubation the samples were all rinsed by  5 successive flows of ethanol and water to remove the extra BPhT molecules, and dried by nitrogen gas. 

On some samples, 1-nm- or 1.5-nm-thick Al$_2$O$_3$ spacers were deposited either on the bare metal, on top of the BPhT SAM or the PVP-capped on AuMP. We used atomic layer deposition (ALD) at 100 $^{\circ}$C. The growth process shows the negligible impact on the Raman signal of the BPhT (Figure S\ref{fig:ALD_impact}), confirming that this temperature is low enough not to damage the molecules and SAM in a significant way.
For the fabrication of a monolayer MoS$_2$ spacer, thin bulk MoS$_2$ was firstly transferred on a metal film by mechanical exfoliation method \cite{castellanos-gomez2014}. The sample was then annealed at 200$^{\circ}$C for 8 hours, forming strong and uniform Au-MoS$_2$ bonding that also restructures the metal surface \cite{chen2018a}. Next, the sample was immersed into acetone solution under ultrasound condition for 3~min to peel off the bulk MoS$_2$, leaving the bottom monolayer (or few-layer) MoS$_2$ on the metal surface.

\subsubsection*{Nanoparticle synthesis and their surface functionalisation} 

The AgNCs were synthesized by the protocol from Ref. \cite{zhou2016}. Briefly, H$_2$O solution of 5 mL 0.02 M CTAC (25 wt\% in H$_2$O, Sigma-Aldrich) and 0.5 mL 0.1 M ascorbic acid (Sigma-Aldrich) were mixed in the glass vial for 10 min preheating at 60 $^{\circ}$C. Then aqueous solutions of CF$_3$COOAg (\SI{50}{\micro\liter}, 10 mM) and FeCl$_3$ (\SI{80}{\micro\liter}, \SI{4.3}{\micro\Molar}) were added to the glass vial. After 3 hours reaction, the products were centrifuged at 14500 rpm and finally stored in aqueous solutions of 0.02 M CTAC.

To fabricate BPhT-covered AuNPs, \SI{200}{\micro\liter} aqueous solution of AuNPs (BBI solutions) with original concentration (optical density 0.88 at 520 nm) were mixed with \SI{200}{\micro\liter} aqueous solution of sodium citrate tribasic dihydrate (10~mM) and \SI{600}{\micro\liter} ethanol solution of BPhT (10~mM) for the replacement. After 2 hours incubation, the products were centrifuged at 14500 rpm and finally stored in H$_2$O.

\subsubsection*{Nanoparticle drop-casting and ALD sealing} 
Plasmonic nanojunctions were formed by drop-casting AuNP solution on various spacer-film systems, where the coverage of the nanoparticles depends on the colloid concentration and the surface condition (hydrophobic vs. hydrophilic). After 30~s to 5~min incubation (depending on solution concentration), the samples were gently rinsed by water and dried by nitrogen gas. 
For AgNCs, the sample was dried after drop-casting, and then immersed into ethanol and H$_2$O to remove the residual CTAC molecules on the AgNCs. 
The CTAC and PVP molecules capped on the NPs and AuMP bring additional 1-2~nm spacer distance in the plasmonic nanocavities \cite{chen2018a}, and $\sim$0.5~nm thickness for citrate capped nanoparticles. 
For  BPhT SAM and MoS$_2$ directly contacting with the ligand layer, the additional gap thickness would be further reduced due to the BPhT replacement and MoS$_2$-induced metal surface migration \cite{chen2018,sigle2015}.
Finally, a 4 to 10~nm compact alumina layer was grown on the sample surface by ALD at 100$^{\circ}$C. This oxide layer protection improves the long-term stability against oxidation and laser irradiation. 
As shown in Supplementary Fig.~\ref{fig:ALD_impact}, the growth of Al$_2$O$_3$ on the sample surface results in the redshift of the plasmonic resonance, due to the combination of the increased charge screening effect and the slightly increased bottom facet size of the nanoparticles under heating during ALD {\cite{chen2018}.

\subsection*{Spectroscopy}

\subsubsection*{Simultaneous PL and Raman measurements}

The simultaneous PL and Raman measurements with 532 nm and 750 nm excitation beams were implemented by the optical setup shown in Supplementary Fig.~\ref{fig:Setup_RP}. 
Linearly polarised 750~nm continuous wave (cw) light from a Ti:Sa laser was sequentially directed to a noise-eater, a cleanup filter and a radial polarisation converter to form a clean and narrow 750 nm laser line with radial polarisation. The laser was eventually focused by a high numerical aperture (NA) objective to form a diffraction-limited laser beam on the sample, providing a large out-of-plane electric field component to effectively excite the gap mode of the nanojunctions. 
On the other hand, linearly polarised 532 nm cw light from a diode laser was directed and collimated though a fiber coupling system, and then pass through a cleanup filter, a radial polarisation converter and the same objective to form a radially polarised 532 nm laser beam with a clean and narrow spectral line on the sample. 
The sample was mounted on an 3-axis piezo-stage with displacement precision better than 100 nm, allowing for three-dimensional alignment of the single nanojunction with respect to the laser beam. The light from the sample was collected by the same objective, then passed through a group of short and long pass filters and a 532 nm notch filter to eliminate 532 nm and 750 nm light, eventually directed to the slit of a spectrometer. A part of the reflected light was directed to a camera to find the position of the nanoparticles by white light illumination and confirm the alignment of the 532 nm and 750 nm laser beams. All the raw spectra were subtracted by noise spectra with the same integration time.

\subsubsection*{Simultaneous PL and DF measurements}

The PL and/or dark-field (DF) measurements were implemented by the optical setups shown in Supplementary Fig.~\ref{fig:PL_DF_S}. 
For DF measurements, white light from a halogen lamp was guided by a multimode fiber, collimated and refocused by two lenses placed on the side of the sample stage, with the beam making an angle of 10$^{\circ}$ with the plane of the sample. A polariser was placed in between the lenses to convert unpolarised white light to linearly polarised light, enabling $p$- or $s$-polarisation excitation. 
The scattered light from the sample (without specular reflection) was collected by an objective lens and  directed to a camera for DF imaging or coupled to a multimode fiber connecting to a spectrometer for DF spectroscopy. The 0.2~mm core size of the fiber gives a circular collection area with a diameter \SI{\sim 2}{\micro\metre} on the sample, allowing for the scattered signal  from only one single nanojunction to enter the spectrometer. A background spectrum was acquired from the bare metal film near the measured nanojunction. The net DF spectrum was then obtained by background subtraction and then divided by the spectrum acquired by directing the white light into the objective (using an angled micro-mirror). This calibration accounts for both the intrinsic lamp spectrum and the setup spectra response. 
The sample was mounted on a 3-axis piezo-stage with displacement precision better than $\sim$100 nm, allowing for precise alignment between the nanoparticle and the collection area. 

For simultaneous PL and DF measurements, a 532~nm laser beam was directed to the objective after a clean-up filter, forming a highly focused beam adjusted to the center region of the collection area. The PL signal was collected in the same way as the DF signal, after blocking wavelengths shorter than 550~nm with a dichroic mirror and a long pass filter. A 532~nm notch filter was placed in front of the camera to image the PL. The PL measurements implemented with the setup in Supplementary Fig.~\ref{fig:PL_DF_S}b follows the same laser excitation and signal collection paths. 
A dark-noise spectrum acquired with the same integration time was subtracted from each raw spectrum.

\subsubsection*{Temperature dependent PL and/or Raman measurements}

The variable-temperature PL and/or Raman measurements were implemented by a cryogenic system shown in Supplementary Fig.~\ref{fig:Cryostat}.
Before the optical measurement, the sample was glued on a 3-axis piezostage (attocube) integrated inside the cryostat by elargol.  The vacuum of the chamber was pumped down to $<5\times10^{-6}$ mbar, following modif{by} cooling down to a base temperature of 3.8~K. The control of the sample temperature above 3.8~K was realized by a heating system integrated on the sample holder, with the precision of 0.1~K. 
For spectroscopy, linearly polarised beams from a Ti:Sa laser tuned to 740 nm, and two other fixed frequency lasers (diode laser at 532~nm and HeNe laser at 633~nm) were overlapped with the help of dichroic mirrors, and reflected on a beam-splitter into the objective.  
A single nanojunction was aligned to the focus position by a piezo-stage with three-dimensional displacement precision better than 10~nm. 
The output light from the sample was collected by the same objective, passed through tunable bandpass filters and notch filters to eliminate 532, 633 nm and 740 nm light. The signal was then either focused into the slit of a spectrometer or into a fiber-coupled to a single photon counting avalanche photodiode (APD)  to measure high-speed intensity traces. 
Part of the reflected light was directed to a camera to confirm the alignment of the laser beams and to find the position of the nanojunctions under white light illumination. 
A dark-noise spectrum acquired with the same integration time was subtracted from each raw spectrum.

\subsection*{{Simulation}}
\label{simulation}
The full-wave simulations were performed on a commercial software package (COMSOL Multiphysics 5.2a). An 80-nm-diameter Au nanoparticle with 40-nm-diameter bottom facet size (truncated sphere) was separated from the Au film by a 1.3- nm-thick layer (refractive index = 1.4). 
The dielectric function of the gold followed experimental data from Johnson \& Christy \cite{johnson1972a}. 
For the simulation shown in the main text, oscillating electric point dipoles were placed on the Au film  (the position from symmetry axis: x = 0~nm, 12~nm, 17~nm) to mimic the PL radiation from localised emitters.
The local enhancement of the photonic density of states vs. frequency was calculated by comparing the dipole radiation with and without the plasmonic geometry. 
A collection cone was applied in the integration to keep only the emission inside a 0.85 numerical aperture. 
%The Purcell spectra reflect that the radiation is highly dependent on the localised EM environment and the spatial mode distribution plays a significant role in the radiation. 
In the experiment, we propose that the randomness of the emitter position makes it couples randomly to different modes. 
Plane-wave excitation was also applied to calculate the basic scattering/absorption spectra (the main text), and clarify the plasmon field and charge distribution.

For the simulation in Supplementary Fig.~\ref{fig:NP150}, a 150-nm-diameter AuNP with 70 nm facet is placed on the silica (refractive index = 1.5) substrate. A 1- nm-thick gap is set between the NP and substrate due to represent the ligand. The scattering is calculated by illuminating a plane wave with an 80 degree incident angle from the surface.

To investigate the laser-induced thermal effect, a 532~nm laser beam with $\sim 3.5\times 10^4~$W/cm$^2$ intensity illuminated the nanojunction. 
The electromagnetic simulation was performed to derive the heat absorption in the metal, which is then plugged into the heat transfer model as a heat source. 
The thermal conductivity of the Au and spacer (ALD, $\textit{k}_\mathrm{Gap}$) was set as 314 W/(m·K) and 1.8 W/(m·K), respectively, while the heat capacity of the two materials was set as 129 J/(kg·K) and 755 J/(kg·K) \cite{young1992,cappella2013}.
%The thermal conductivity of the Au and spacer was set as 341 W/(m·K) and 2.5 W/(m·K), respectively, while the heat capacity of the two materials was set as 129 J/(kg·K) and 880 J/(kg·K). 
%The accumulated heat would be easily transferred into the bath.

\clearpage
%\subsection*{List of investigated nanojunctions}
%-------------------------------------------------------------------------------------------------------------------------------------------------------------------------------------------------------
\begin{table}[htp!]
    \centering
    \includegraphics[width=1\columnwidth]{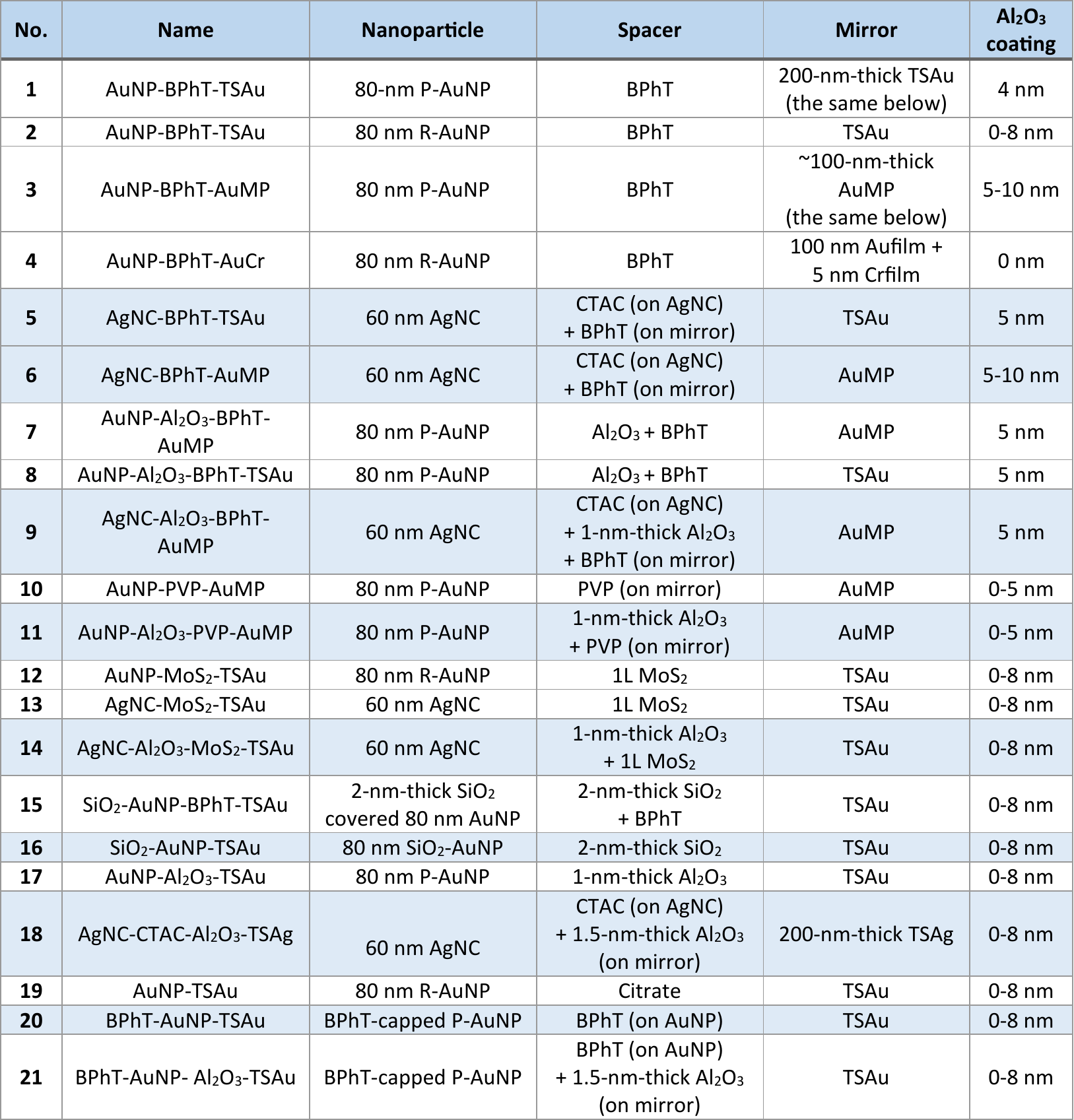}
    \caption{
    \textbf{List of plasmonic nanojunctions investigated.} 
    P-AuNP and R-AuNP (see Supplementary Fig.~\ref{fig:SEMDF}a): polyhedral and rounded Au nanoparticle, AuMP: Au microplate, AgNC (see Supplementary Fig.~\ref{fig:SEMDF}b): Ag nanocube, TSAu and TSAg: template-stripped Au See Supplementary Fig.~\ref{fig:AFM}a) and Ag film.
    }
\label{tab:Sample_list}
\end{table}
%-------------------------------------------------------------------------------------------------------------------------------------------------------------------------------------------------------

\clearpage
%\subsection*{SEM and optical image of nanoparticles and nanojunctions}
%-------------------------------------------------------------------------------------------------------------------------------------------------------------------------------------------------------
\begin{figure}[htp]
    \centering
    \includegraphics[width=0.6\columnwidth]{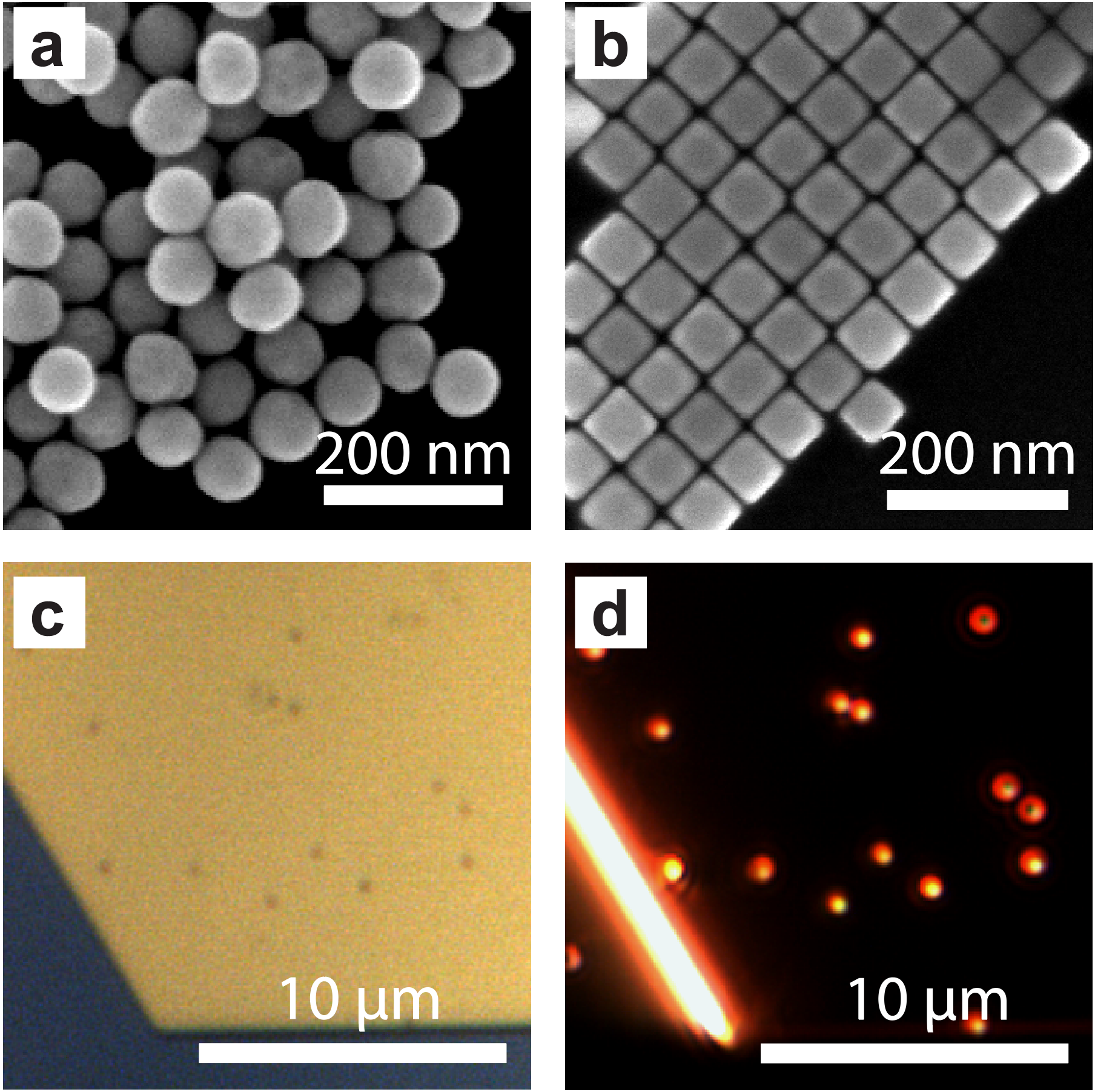}

    \caption{
    \textbf{SEM and optical image of nanoparticles and nanojunctions.}
(\textbf{a, b}) SEM top view of (\textbf{a}) gold nanospheres with $\sim$80 nm diameter (Expedeon) and (\textbf{b}) silver nanocubes with $\sim$ 60 nm edge size (home made). (\textbf{c}) Top view of bright field and (\textbf{d}) DF scattering images of AuNP-BPhT-AuMP nanojunctions (No.3 in Supplementary Table \ref{tab:Sample_list}). 
Nanoparticles can be well located even under bright field illumination as darker spots.
}
    \label{fig:SEMDF}
\end{figure}
%-------------------------------------------------------------------------------------------------------------------------------------------------------------------------------------------------------

\clearpage
%\subsection*{AFM images of Au film surfaces}
%-------------------------------------------------------------------------------------------------------------------------------------------------------------------------------------------------------
\begin{figure}[htp]
    \centering
    \includegraphics[width=0.9\columnwidth]{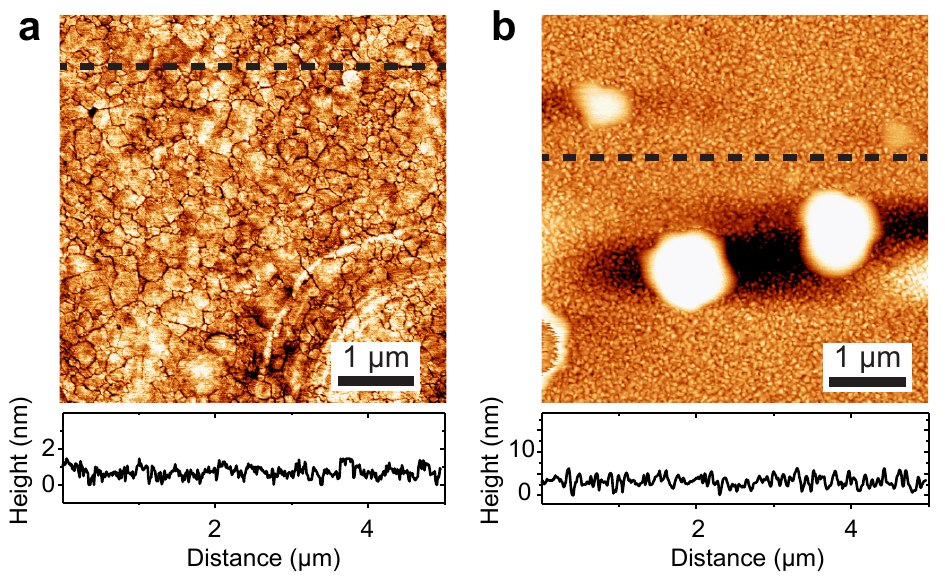}
    \caption{
    \textbf{AFM images of Au film surfaces.}
(\textbf{a, b}) Representative atomic force microscope (AFM) images of TSAu (\textbf{a}) and AuCr (\textbf{b}) films, with height profiles along the lines superimposed on each image shown in the bottom panel. 
Based on grain size analysis we find that the mean grain size of TSAu and AuCr are 154$\pm$19 nm and 45$\pm$21 nm, respectively.
Bright and irregular features on AuCr film (\textbf{b}) are most likely contaminants. 
}

    \label{fig:AFM}
\end{figure}
%-------------------------------------------------------------------------------------------------------------------------------------------------------------------------------------------------------

\clearpage
%\subsection*{Impact of atomic layer deposition}
%-------------------------------------------------------------------------------------------------------------------------------------------------------------------------------------------------------

\begin{figure}[htp]
    \centering
    \includegraphics[width=0.5\columnwidth]{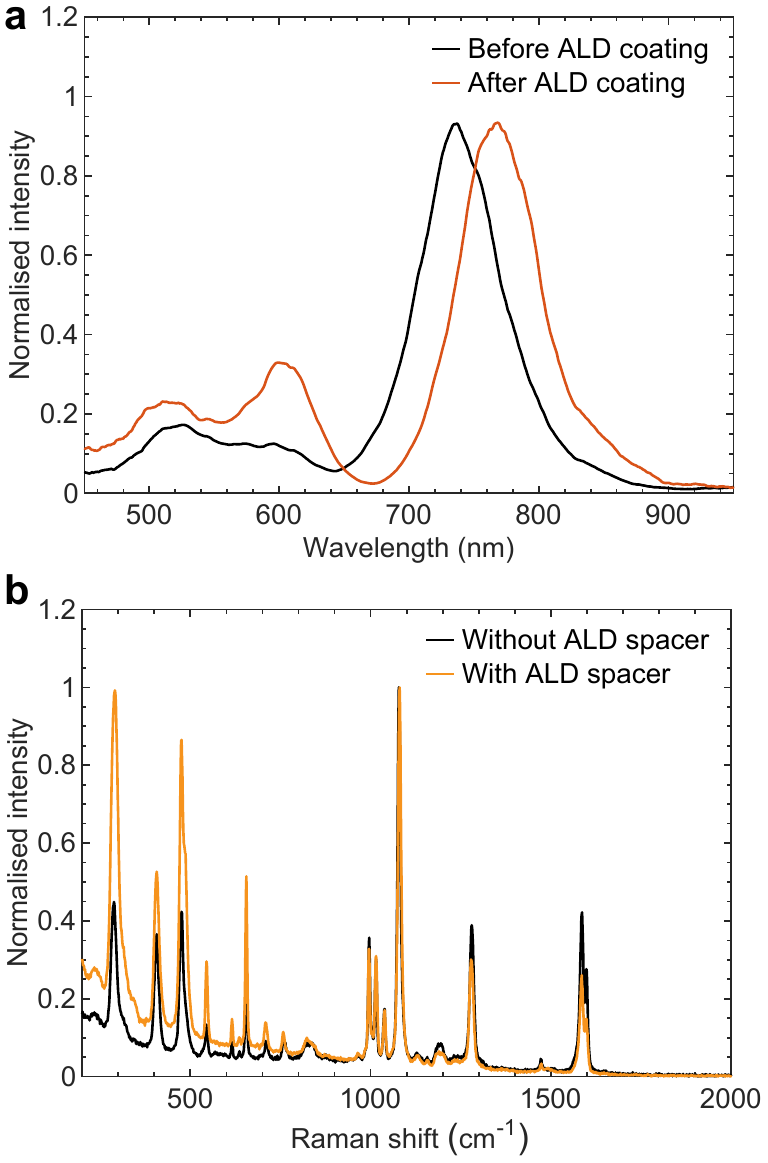}
    \caption{\textbf{Impact of ALD used for spacer or capping layer.}
    (\textbf{a}) DF scattering spectra of a plasmonic nanojunction before and after Al$_2$O$_3$ capping of the sample. The redshift is expected due to the increase in the dielectric constant around the particle. 
    (\textbf{b}) Raman spectra of the BPhT plasmonic nanojunction with (No.~3 in Supplementary Table~\ref{tab:Sample_list}) and without (No.~7 in Supplementary Table~\ref{tab:Sample_list}) Al$_2$O$_3$ layer as an additional spacer.
    The slightly increased Raman signal at small Raman shift relative to large Raman shift when the additional ALD spacer is inserted may be due to a shift of the plasmonic resonances (blue shift because of a larger gap) closer to the laser wavelength of 785~nm in this case. 
}
    \label{fig:ALD_impact}
\end{figure}
% Supplementary Fig.~\ref{fig:PL_Raman}B    ($\sim 5$~nm)  \textit{not} fluctuating 
%-------------------------------------------------------------------------------------------------------------------------------------------------------------------------------------------------------

%1111111111111111111111111111111111111111111111111111111111111111111111111111111111111111111111111111111111111111111111111111111

%2222222222222222222222222222222222222222222222222222222222222222222222222222222222222222222222222222222222222222222222222222222
\clearpage
%\section{Experimental setup}

%\input{setup}
%\subsection*{Schematic of the optical setup for PL+Raman}
%-------------------------------------------------------------------------------------------------------------------------------------------------------------------------------------------------------
\begin{figure}[htp!]
    \centering
    \includegraphics[width=0.9\columnwidth]{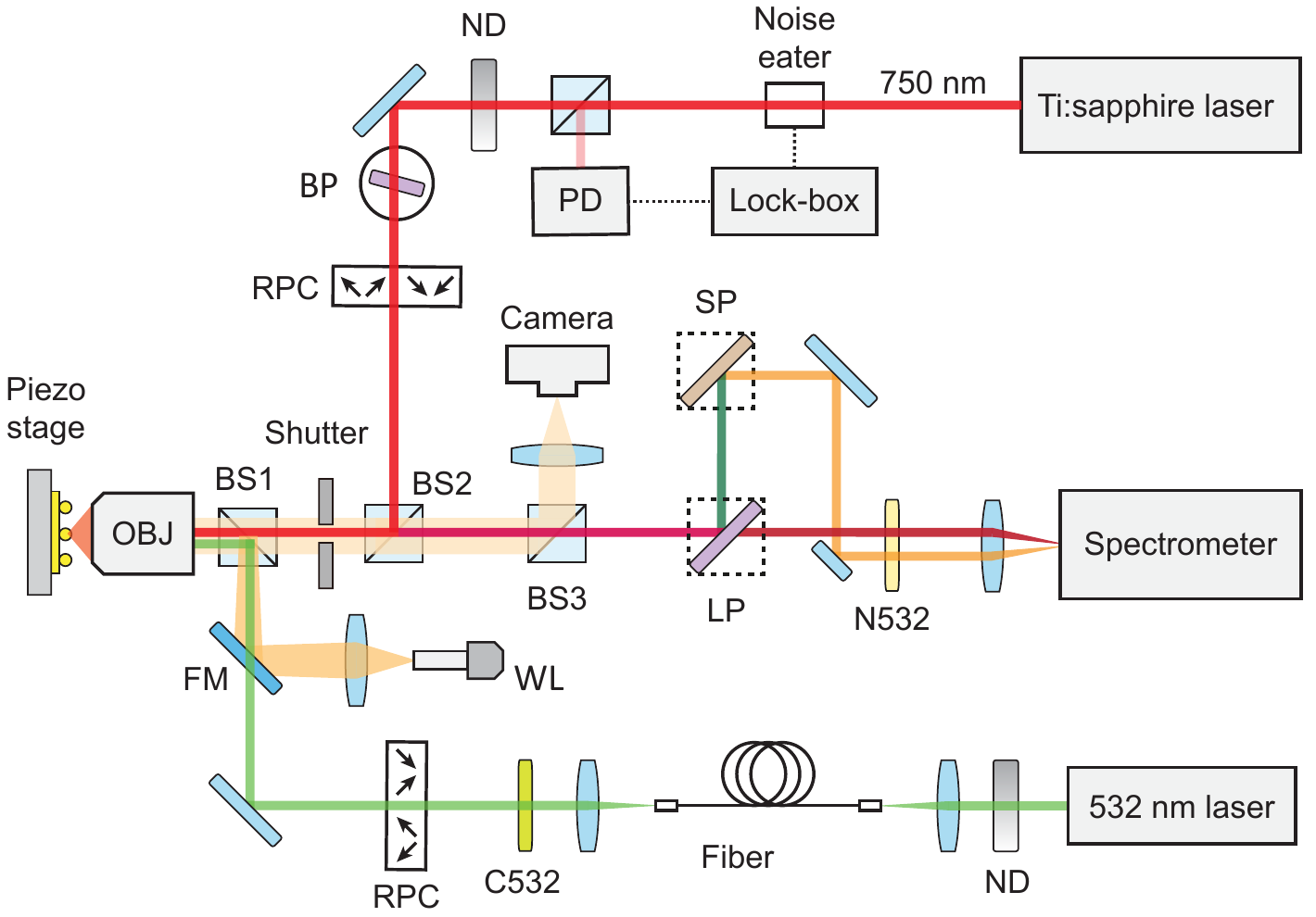}
    \caption{
     \textbf{Schematic of the light path for simultaneous PL and Raman measurement.}
 ND: neutral density filter, FM: flip mirror, BP: tunable bandpass filter, RPC: radial polarisation converters, PD: photodetector, WL: white light source, BS1: beamsplitter with reflection:transmission in \% (R:T) = 10:90, BS2: beamsplitter with R:T = 20:80, BS3: flip pellicle beamsplitter with R:T = 8:92, LP: AHF tunable longpass filter module, SP: AHF tunable shortpass filter module, N532: notch filter centered at 532 nm, C532: cleanup filter centered at 532 nm, OBJ: Nikon objective, numerical aperture = 0.95, working distance = 0.21 mm, Spectrometer: Andor Shamrock, grating: 300 l/mm, CCD: Andor iDus 416, Tunable Ti:Sa laser: Coherent MBR, tuned to 750 nm. 3-axis stage: The sample is mounted on an 3-axis piezo stage with displacement precision better than 100 nm. 
}
    \label{fig:Setup_RP}
\end{figure}
%-------------------------------------------------------------------------------------------------------------------------------------------------------------------------------------------------------

\clearpage
%\subsection*{Schematic of the optical setups for PL and dark-field (DF) spectroscopy}
%-------------------------------------------------------------------------------------------------------------------------------------------------------------------------------------------------------
\begin{figure}[htp!]
    \centering
    \includegraphics[width=0.9\columnwidth]{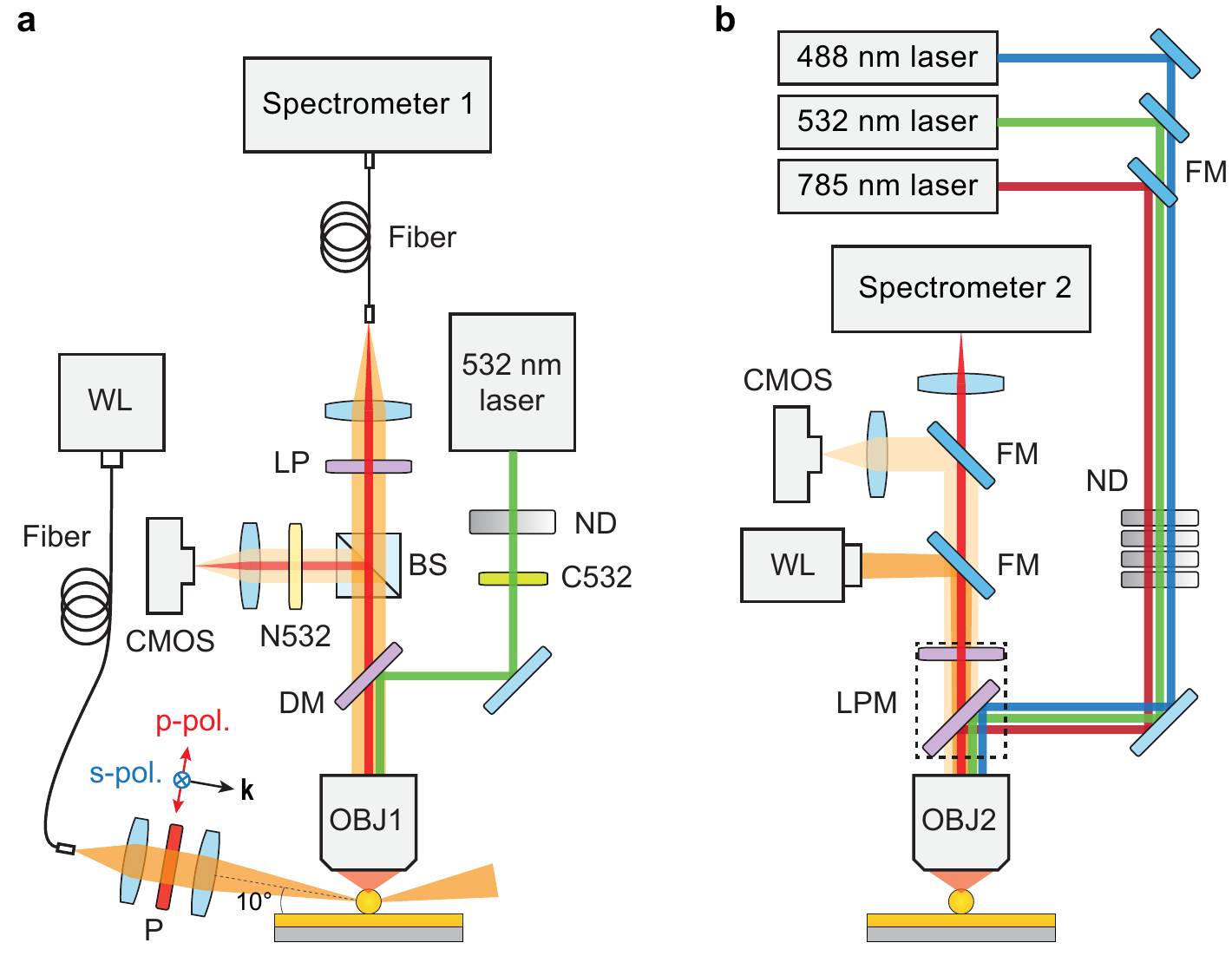}
    \caption{
    \textbf{Schematic of the optical setups for PL and dark-field (DF) spectroscopy}
 (\textbf{a}) ND: neutral density filter. C532: cleanup filter at 532 nm. P: broadband polariser. DM: dichroic mirror. N532: notch filter centered at 532 nm. LP: 540 nm longpass filter. BS: beamsplitter with R:T = 30\%:70\%. FM: beamsplitter with R:T = 10\%:90\%. WL: white light source. OBJ1: objective (Olympus), numerical aperture = 0.8, working distance = 3.1 mm. Spectrometer1: QEpro, Ocean Optics with a grating of 100 lines/mm, coupled via a 0.1 mm core size multimode fiber.  
 (\textbf{b}) Spectrometer2: Renishaw inVia, 300 l/mm for 488 nm and 532 nm laser and 1200 l/mm grating for 785 nm laser. LPM: switchable longpass filter module. FM: flip mirror. OBJ2: objective (Leica), numerical aperture = 0.85, working distance = 1 mm. 
}
    \label{fig:PL_DF_S}
\end{figure}
%-------------------------------------------------------------------------------------------------------------------------------------------------------------------------------------------------------

\clearpage
%\subsection*{Schematic of the cryostat optical setup}
%-------------------------------------------------------------------------------------------------------------------------------------------------------------------------------------------------------
\begin{figure}[htp!]
    \centering
    \includegraphics[width=0.9\columnwidth]{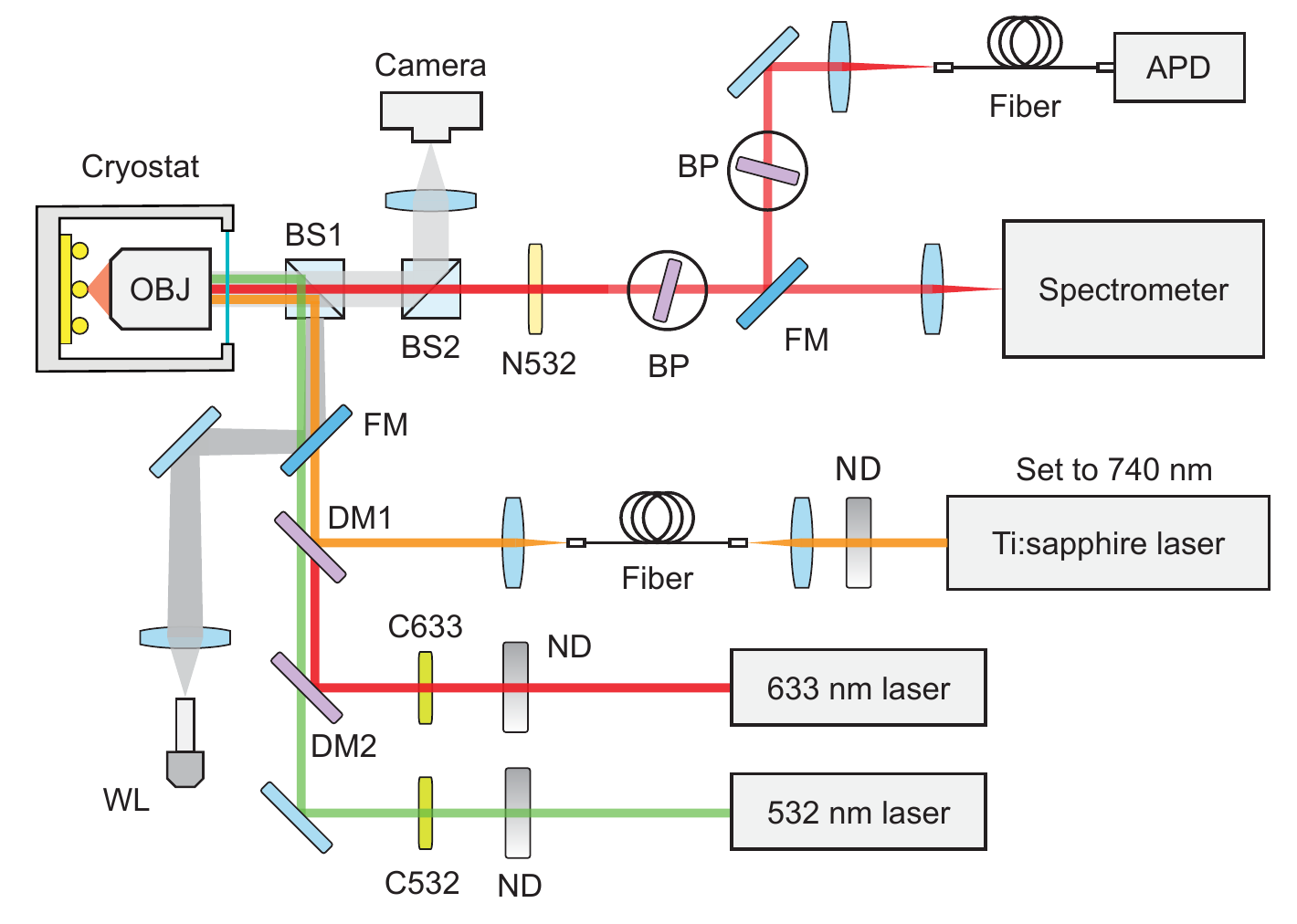}
    \caption{
    \textbf{Schematic of the light path for PL +Raman measurement at 3.8-300 K.}
 ND: neutral density filter. BS1: beamsplitter R:T = 8\%:92\%. BS2: beamsplitter with R:T = 10\%:90\%. C532 and C633: cleanup filter at 532 nm and 633 nm. N532: notch filter at 532 nm. BP: tunable bandpass filter module covering 550-950 nm with bandwidth $\sim$100 nm. WL: supercontinuum source used for bright field imaging. FM: flip mirror. DM1: dichroic mirror. DM2: dichroic mirror. Ti:sapphire laser: Spectra-Physics 3900S, tuned to 740 nm.
Cryostat: closed-cycle (cyostat attoDRY800, Attocube), the objective and piezo stage are integrated inside the chamber. OBJ: objective, numerical aperture = 0.81, working distance = 0.5 mm. Spectrometer: Andor Kymera 193i, grating: 600 l/mm. APD: a single-mode fiber-coupled avalanche photodiode, Excelitas Technologies.
    }
    \label{fig:Cryostat}
\end{figure}
%-------------------------------------------------------------------------------------------------------------------------------------------------------------------------------------------------------

\clearpage
%\subsection*{Measurement noise}
%-------------------------------------------------------------------------------------------------------------------------------------------------------------------------------------------------------

\begin{figure}[htp!]
    \centering
    \includegraphics[width=0.7\columnwidth]{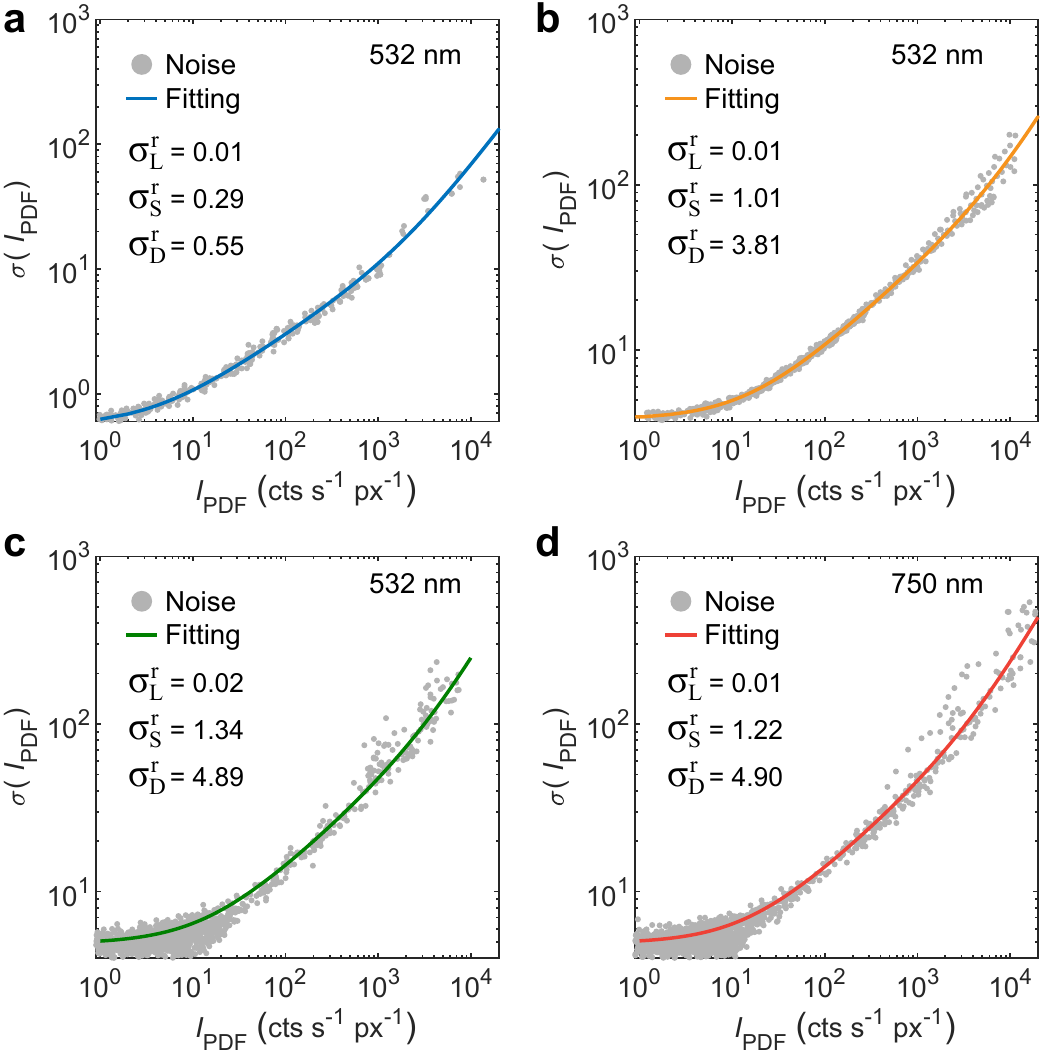}
    \caption{\textbf{Analysis of measurement noise.}
    Measurement noise of the setup used (\textbf{a}) in Supplementary Fig.~\ref{fig:PL_DF_S}b and (\textbf{b}) in Supplementary Fig.~\ref{fig:Cryostat} (\textbf{c} and \textbf{d}) in Supplementary Fig.~\ref{fig:Setup_RP}, all of which were used to acquire the blinking traces reported in the main text. 
    These data were acquired on the laser beam attenuated to obtain count rates matching the range of PL and/or Raman count rates typically observed. 
For a given mean signal intensity $I$ the standard deviation of the intrinsic measurement noise $\sigma(I)$ has contributions from classical laser intensity fluctuations $\sigma_\mathrm{L}=\sigma^\mathrm{r}_\mathrm{L}I$, scaling linearly with $I$, from photon shot noise $\sigma_\mathrm{S}=\sigma^\mathrm{r}_\mathrm{S}\sqrt{I}$, and from detector noise $\sigma_\mathrm{D}$ (mainly readout noise of the CCD; for the short acquisition times used here modif{the} dark current is negligible), according to 
$\sigma(I)=\sqrt{{(\sigma^\mathrm{r}_\mathrm{L}I)}^2+{(\sigma^\mathrm{r}_\mathrm{S}\sqrt{I})}^2+{\sigma_\mathrm{D}}^2}$ \cite{homola2006}. 
This formula is used to fit the raw data, with the parameters shown in the figure.
The shaded blue areas in the probability density function (PDF) in the main text and supplementary materials follow a Gaussian distribution given by 
$f_\mathrm{PDF}(I) = \exp{[{-(I-I_\mathrm{PDF})^2} /{2{[\sigma(I_\mathrm{PDF})]}^2}]}$, where $I_\mathrm{PDF}$ is the maximum of the PDF of the intensity trace to be calculated.
In the time traces the blue areas are defined as 
$I_\mathrm{PDF} - {3\sigma(I_\mathrm{PDF})} \leq I \leq I_\mathrm{PDF} + {3\sigma(I_\mathrm{PDF})}$, 
where $I_\mathrm{PDF}$ is used to present the average value of the baseline PL.
    } 
    \label{fig:noise}
\end{figure}
%-------------------------------------------------------------------------------------------------------------------------------------------------------------------------------------------------------

%2222222222222222222222222222222222222222222222222222222222222222222222222222222222222222222222222222222222222222222222222222222

%3333333333333333333333333333333333333333333333333333333333333333333333333333333333333333333333333333333333333333333333333333333
\clearpage
%\section{Investigation of the blinking mechanism}

%\subsection*{\textcolor{red}{Theoretical description of the PL process}}

%\subsection*{Normalisation of PL spectra}
%-------------------------------------------------------------------------------------------------------------------------------------------------------------------------------------------------------
\begin{figure}[h!]
    \centering
    \includegraphics[width=1\columnwidth]{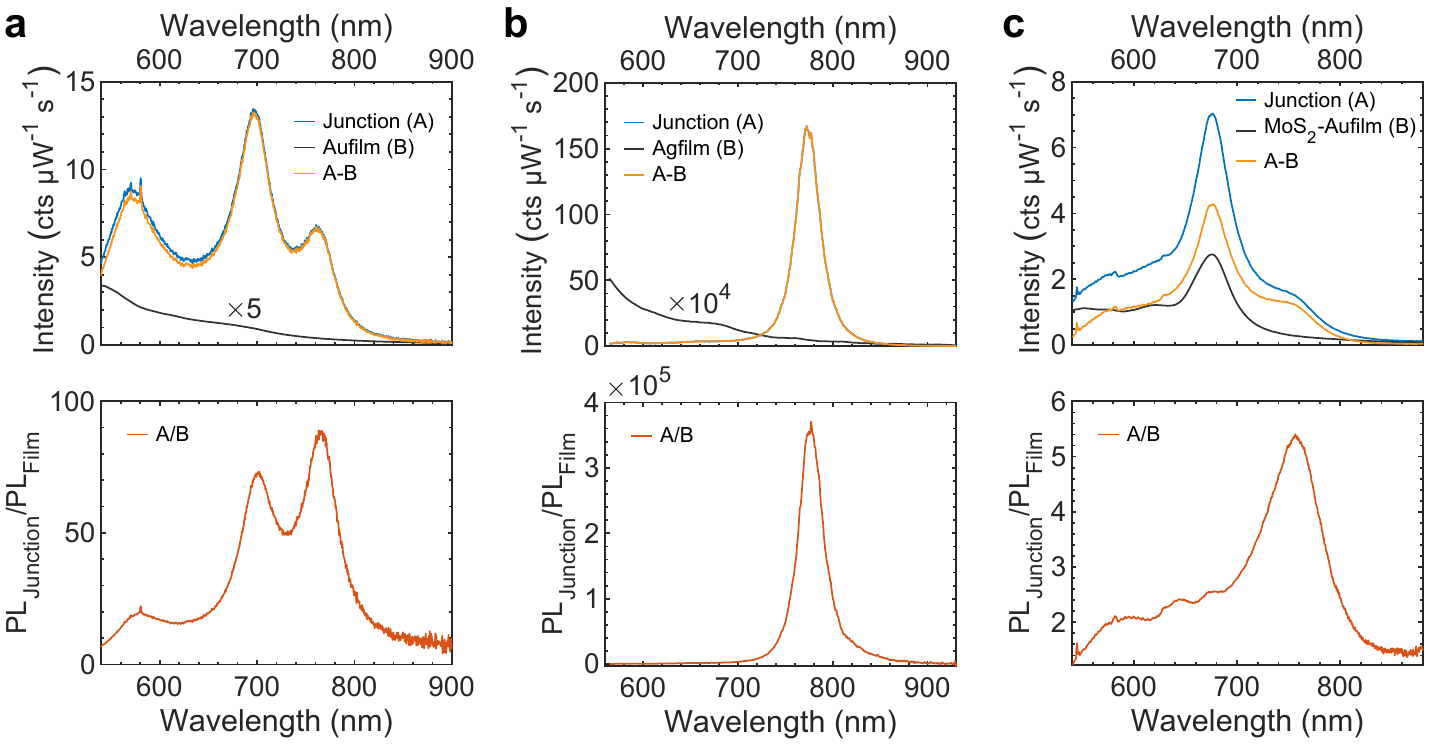}
    \caption{
    \textbf{Normalisation of PL spectra.}
    (\textbf{a})  (top panel) Example of time-averaged PL spectrum when the objective focus is centered on a Au nanojunction (labeled A), or on an area of Au film without nanoparticle (magnified 5 times for visibility, labeled B). 
    The PL spectra shown in this article are obtained by subtracting the latter from the former (A-B).
    The nanojunction PL enhancement factor can be estimated by dividing the nanojunction PL by the film PL:  (A-B)/B$\simeq$A/B (the last approximation is valid because B$\ll$A). But this formula does not account for the much smaller area of the nanojunction compared with the spot size, and thus yields enhancement factors that are about 3 orders of magnitude underestimated  \cite{lumdee2014,cai2018}. 
    (\textbf{b}) The corresponding spectra for an Ag nanojunction and the bare Ag film and their subtraction and division results. The film emission had to be magnified $10^4$ times to be visible on the same scale.
    (\textbf{c}) PL spectra of a monolayer MoS$_2$ spaced Au nanojunction and the MoS$_2$ reference spectrum on a gold film (bottom panel). 
The excitonic emission of monolayer MoS$_2$ is largely quenched due to metal doping \cite{chen2018,mcdonnell2014} but is still visible and stable.
Using PL of the MoS$_2$ instead of the bare film spectrum as the reference enables us to extract the plasmonic response from the A-exciton background of MoS$_2$ (lower panel), which shows a metal-related PL peak at longer wavelengths. In this particular case, we cannot exclude that a dark exciton is rendered brighter by the Purcell effect and contributes to the new plasmonic emission.
    }
    \label{fig:PL_norm}
\end{figure}
%-------------------------------------------------------------------------------------------------------------------------------------------------------------------------------------------------------

\clearpage

%\subsection*{Comparison of PL and polarisation-dependent DF spectra}

\begin{figure}[htp!]
    \centering
    \includegraphics[width=0.9\columnwidth]{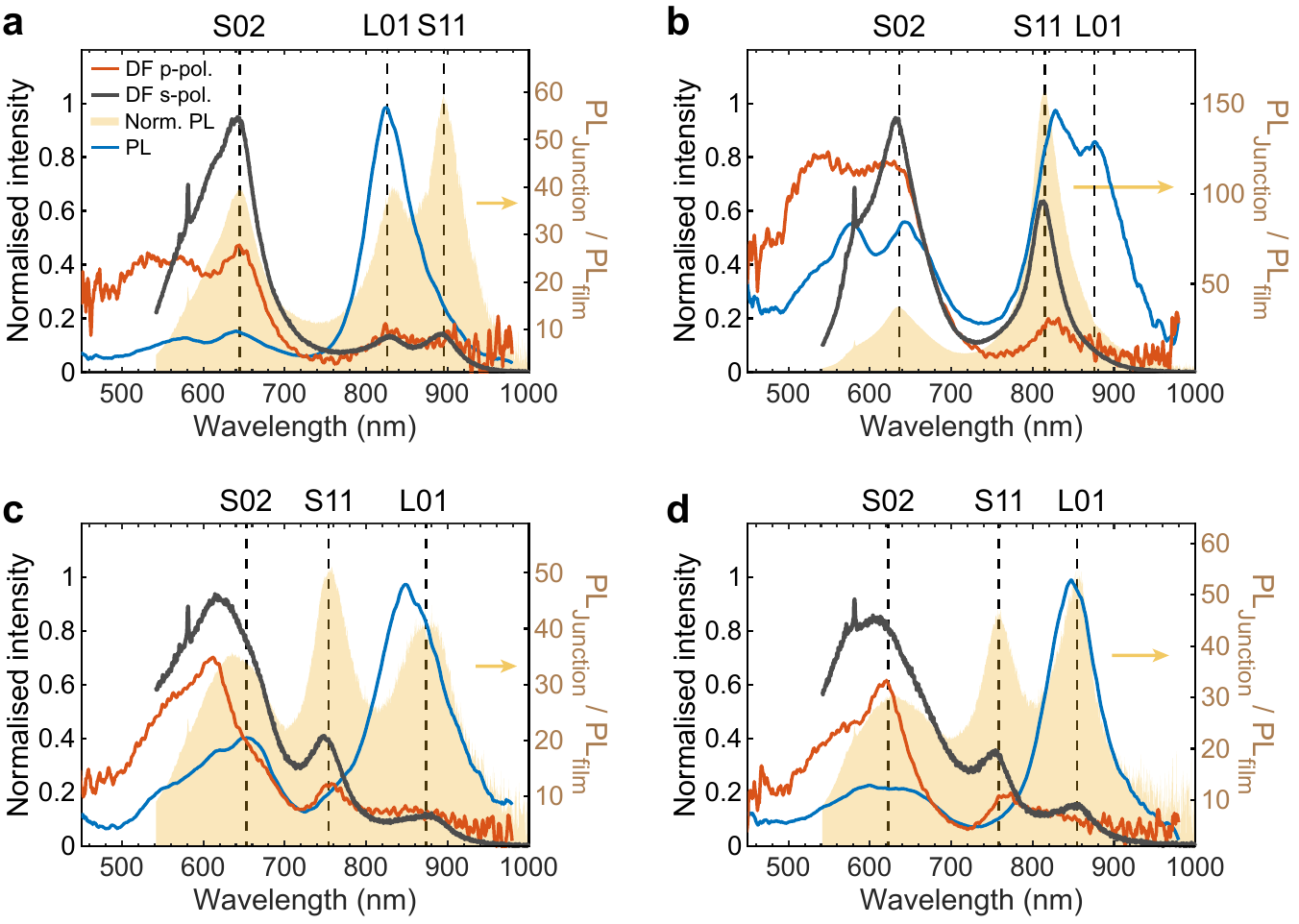}
    \caption{
    \textbf{Comparison of PL and polarisation-dependent DF spectra}
DF scattering spectra of 4 nanojunctions (No. 3 in Supplementary Table~\ref{tab:Sample_list}) excited by s- and p- polarised white light (see the detailed setting in Supplementary Fig.~\ref{fig:PL_DF_S}), along with their normalized time-averaged PL spectra under 532~nm excitation. 
Under p-polarised excitation, L01, S11 and S02 modes can all be excited, with scattering from the L01 antenna mode generally dominating the spectrum. 
With s-polarised excitation, scattering from S11 and S02 can still be detected with similar efficiency, while the L01 mode is much less visible, sometimes invisible. Our results agree well with our simulation analysis and previous reports \cite{chen2018}.
Interestingly, the PL spectral profile typically matches better that of the s-polarised DF than that of p-polarised DF. 
In particular, the S11 mode seems to be excited more efficiently than the L01 mode. 
The origin of this feature could be associated with the better spatial overlap of the near field distribution of the S11 mode with that of the transverse mode (resonant with PL excitation, see Supplementary Fig.~\ref{fig:Heating_sim2}). 
After dividing PL$_\mathrm{Junction}$ by PL$_\mathrm{Film}$ (shaded areas) to eliminate the spectral variations related to the electronic structure of the metal, we find a qualitative match between the normalized time-averaged PL and the DF spectra.
}
    \label{fig:DF_PSpol}
\end{figure}
%-------------------------------------------------------------------------------------------------------------------------------------------------------------------------------------------------------

%%%%%%%%%%%%%%%%%%%%%%%%%%%%%%%%%%%%%%%%%%%%%%%%%%%%%%%%%%%%%%%%%%%%%%%%%%%%
\begin{figure*}[htp]
    \centering
   \includegraphics[width=1.0\columnwidth]{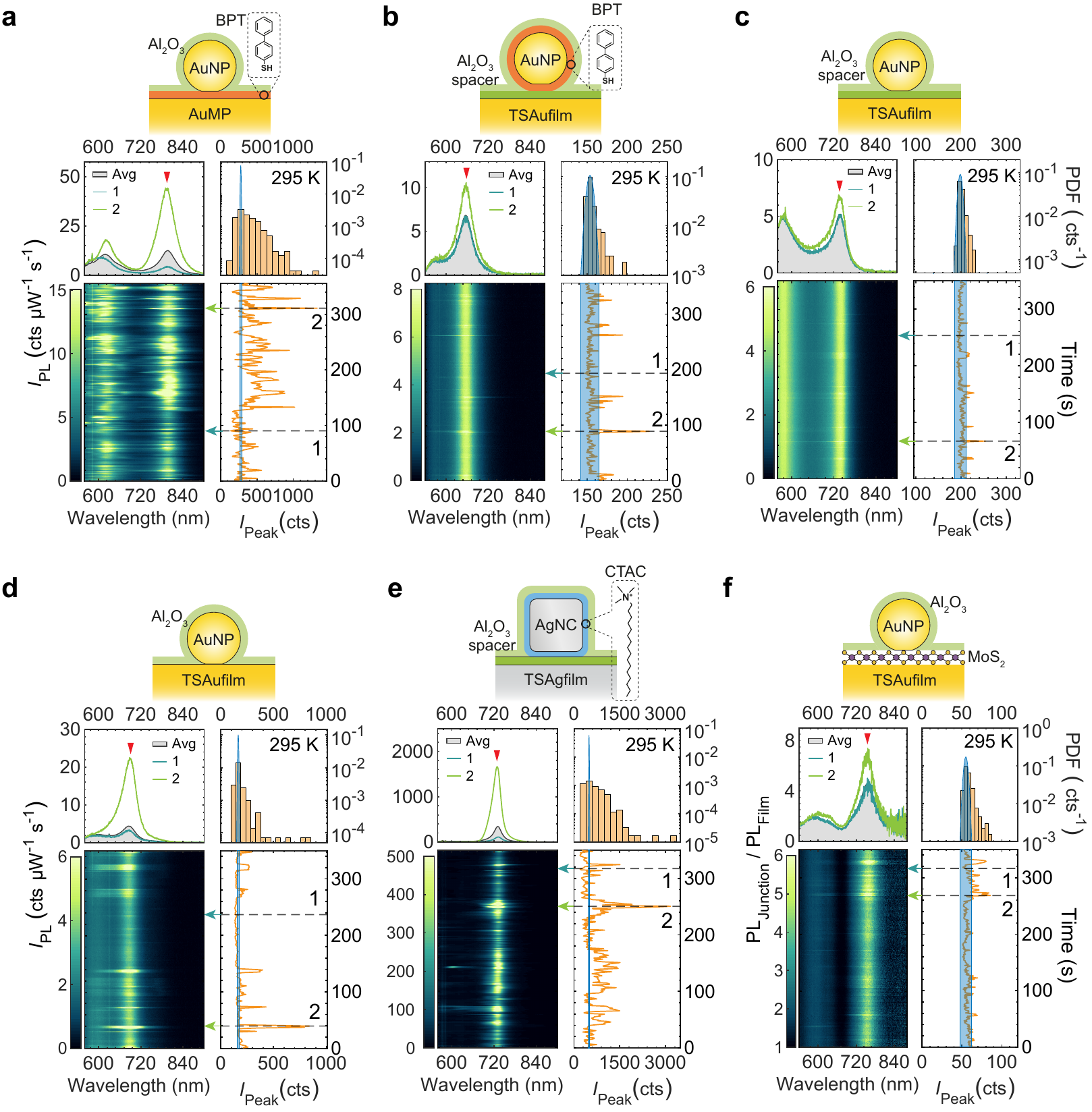}
    \caption{
    \textbf{Comparison of plasmon-enhanced PL blinking from different nanojunction assemblies.} 
    Representative time traces of plasmonic PL emission under continuous 532~nm excitation (full spectra as color plot and peak intensity as the orange line), with examples of single spectra (top-left panel) and probability density function (PDF) for the peak intensity (top-right panel). The blue shaded areas show the irreducible measurement noise for the same count rate. NP: nanoparticle; MP: micro-plates; BPhT: biphenyl thiol; TS: template-stripped; CTAC: cetyltrimethylammonium chloride. TSAufilm: template-stripping Au film. AuMP: Au micro-flake.
    %Alumina layers (5 to 8~nm thick) were created by atomic layer deposition. 
Exposure time is 1s for all panels. The 532~nm laser was focused through a 0.85 NA objective with \SI{37}{\micro\watt} incident power (power density 
$\sim 7.5 \times 10^3$ \SI{}{\watt\per\square\centi\metre}
%$\sim 7.5\times 10^3~$W/cm$^2$
) -- except for panel \textbf{a} (\SI{30}{\micro\watt}) and \textbf{f} (\SI{19}{\micro\watt}).
    }
    \label{fig:comparison}
\end{figure*}

%\subsection*{{PL blinking from additional types of samples}}
%-------------------------------------------------------------------------------------------------------------------------------------------------------------------------------------------------------
\begin{figure}[htp!]
    \centering
    \includegraphics[width=1\columnwidth]{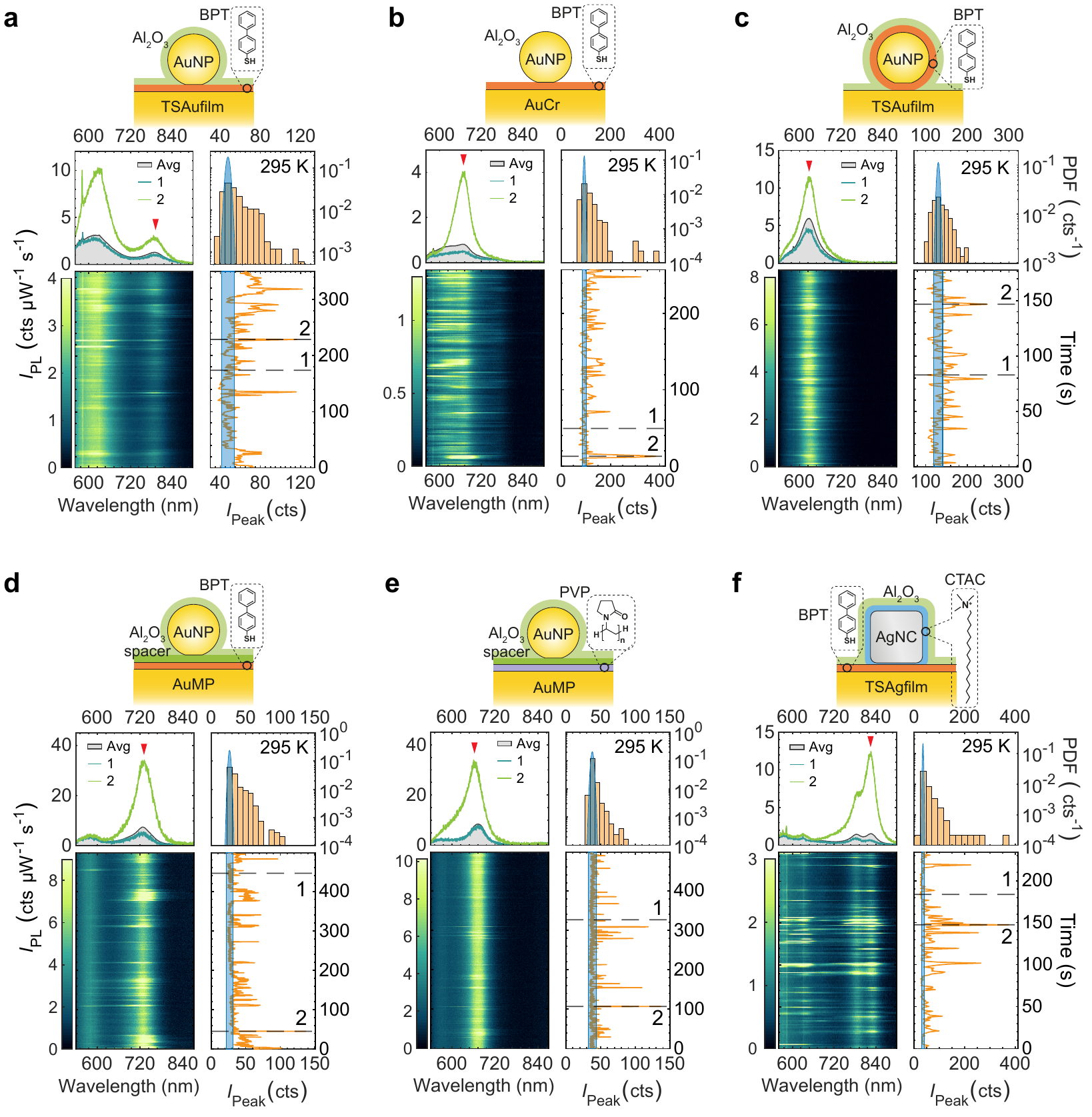}
    \caption{
    \textbf{More nanojunction examples of PL blinking.}
Time series of PL under 532 nm excitation at room temperature from 
(\textbf{a}) a AuNP-BPhT-TSAu nanojunction (from sample No.~1 in Supplementary Table \ref{tab:Sample_list}), 
(\textbf{b}) a AuNP-BPhT-AuCr nanojunction (from sample No.~4), 
(\textbf{c}) a BPhT-AuNP-TSAu nanojunction (from sample No.~20), 
(\textbf{d}) a AuNP-Al$_2$O$_3$-BPhT-AuMP nanojunction (from sample No.~7), 
(\textbf{e}) a AuNP-Al$_2$O$_3$-PVP-AuMP nanojunction (from sample No.~11), 
and (\textbf{f}) AgNC-BPhT-TSAu nanojunction (from  sample No.~5).
The laser intensity was adapted to give well-resolved signal in each case, and remains in the range of few tens of microwatts or less.
%\end{small}  
    }

    \label{fig:more_type}
\end{figure}
%-------------------------------------------------------------------------------------------------------------------------------------------------------------------------------------------------------

\begin{figure}[h!]
    \centering
    \includegraphics[width=0.8\columnwidth]{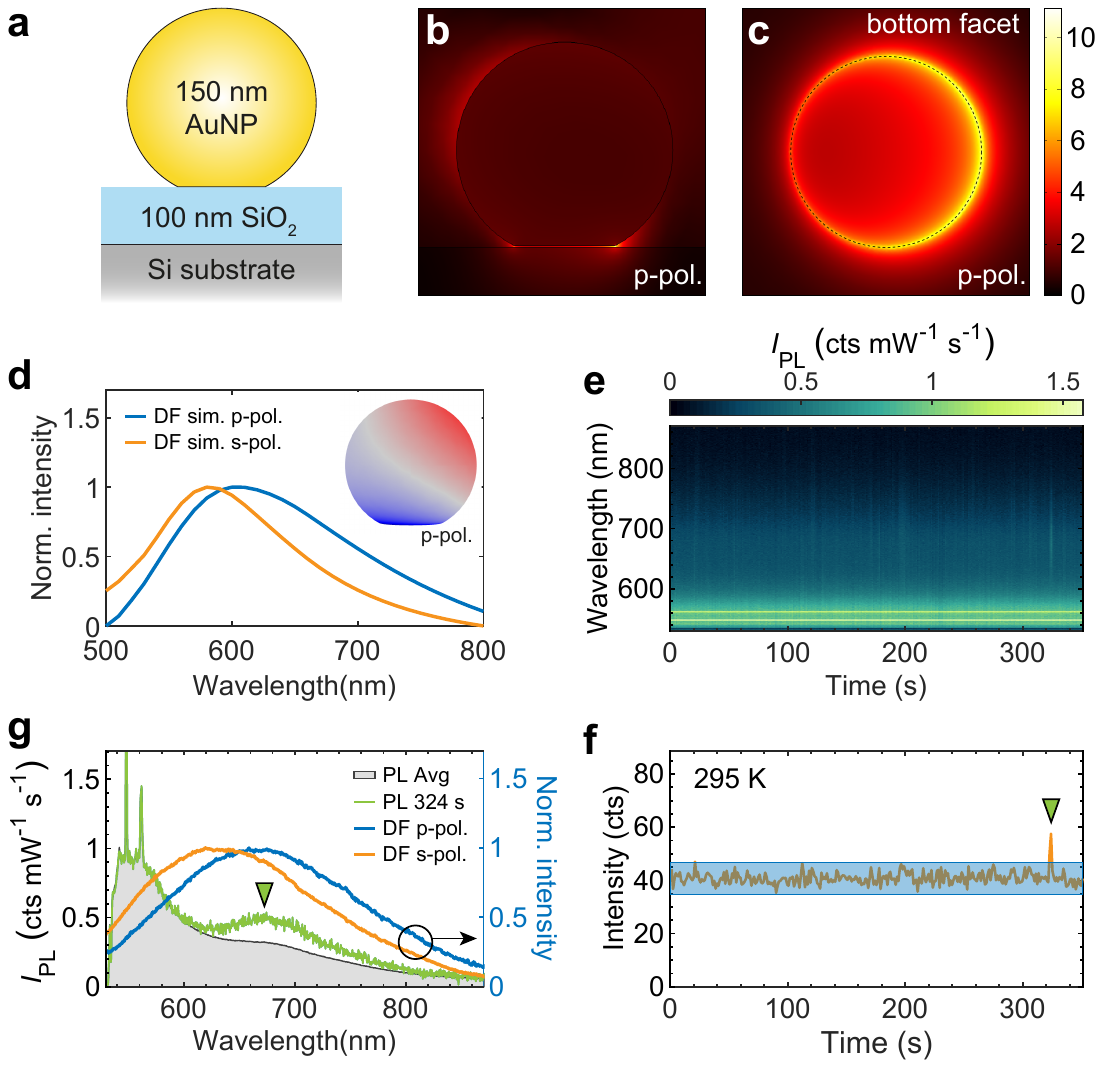}
    \caption{
\textbf{Investigation of an individual 150-nm-diameter Au nanoparticle.}
(\textbf{a}) Schematic of a 150 nm Au nanoparticle on 100-nm-thick SiO$_2$ over a Si substrate.
(\textbf{b}) Vertical cross-section and (\textbf{c}) in-plane bottom facet view of simulated near field enhancement under p-polarised excitation with a plane wave ($k$-vector making a 10$^\circ$ angle with substrate). Details of simulation settings are presented in Sec.~\ref{simulation}.    
(\textbf{d}) Simulated far-field scattering spectra under p- and s-polarised (defined in Supplementary Fig.~\ref{fig:PL_DF_S} and Fig. 1b in the main text) white light illumination. The inset shows the charge distribution at 600~nm under p-polarised excitation, identified as a substrate modified dipole mode. 
(\textbf{e}) An intensity map of PL time series from a Au nanoparticle under 532~nm excitation. Laser power, \SI{110}{\micro\watt}; exposure time, 1s.
(\textbf{f}) The intensity trace of the PL maximum at the dipole mode compared with the measurement noise.
(\textbf{g}) Measured DF spectra under p- and s-polarised white light illumination, and PL spectra of the blinking event and of the time-average from (\textbf{e}). The two peaks around 550 nm are first- (saturated in the figure) and second-order Raman peaks of Si.
    }
    \label{fig:NP150}
\end{figure}
%-------------------------------------------------------------------------------------------------------------------------------------------------------------------------------------------------------

\clearpage
%\subsection*{Wavelength-dependent PL measurements}
%-------------------------------------------------------------------------------------------------------------------------------------------------------------------------------------------------------
\begin{figure}[htp!]
    \centering
    \includegraphics[width=0.9\columnwidth]{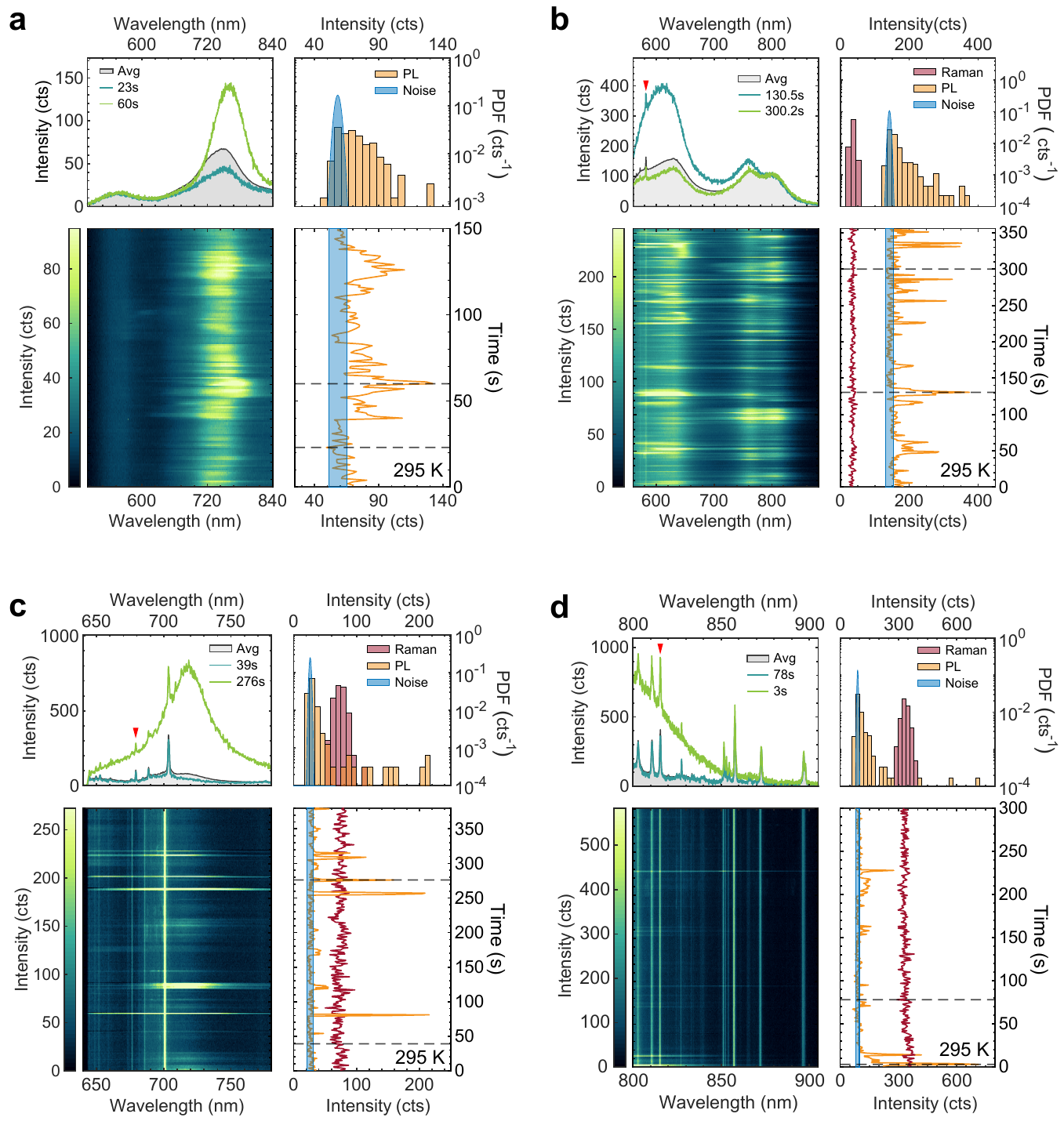}
    \caption{
    \textbf{Wavelength-dependent PL measurements.}
     Time series of PL (and Raman) spectra from nanojunctions as a function of excitation wavelength.
(\textbf{a}) Sample No.~3 in Supplementary Table \ref{tab:Sample_list} with 488 nm excitation with a power density of $\sim1\times10^{4}$ W/cm$^{2}$. 
(\textbf{b}) Sample No.~3 with 532 nm excitation with a power density of $\sim3\times10^{3}$ W/cm$^{2}$. .
(\textbf{c}) Sample No.~18 with 633 nm excitation with a power density of $\sim8\times10^{3}$ W/cm$^{2}$. 
(\textbf{d}) Sample No.~3 with 785 nm excitation with a power density of $\sim2\times10^{4}$ W/cm$^{2}$. 
Note that under 785 nm excitation only the intraband transition and electronic Raman scattering are accessible. Exposure time are all 1 s.
}
    \label{fig:PL_from488to785}
\end{figure}

%-------------------------------------------------------------------------------------------------------------------------------------------------------------------------------------------------------

\clearpage
%\subsection*{PL + Raman at cryogenic temperature}
%-------------------------------------------------------------------------------------------------------------------------------------------------------------------------------------------------------

\begin{figure}[htp!]
    \centering
    \includegraphics[width=0.9\columnwidth]{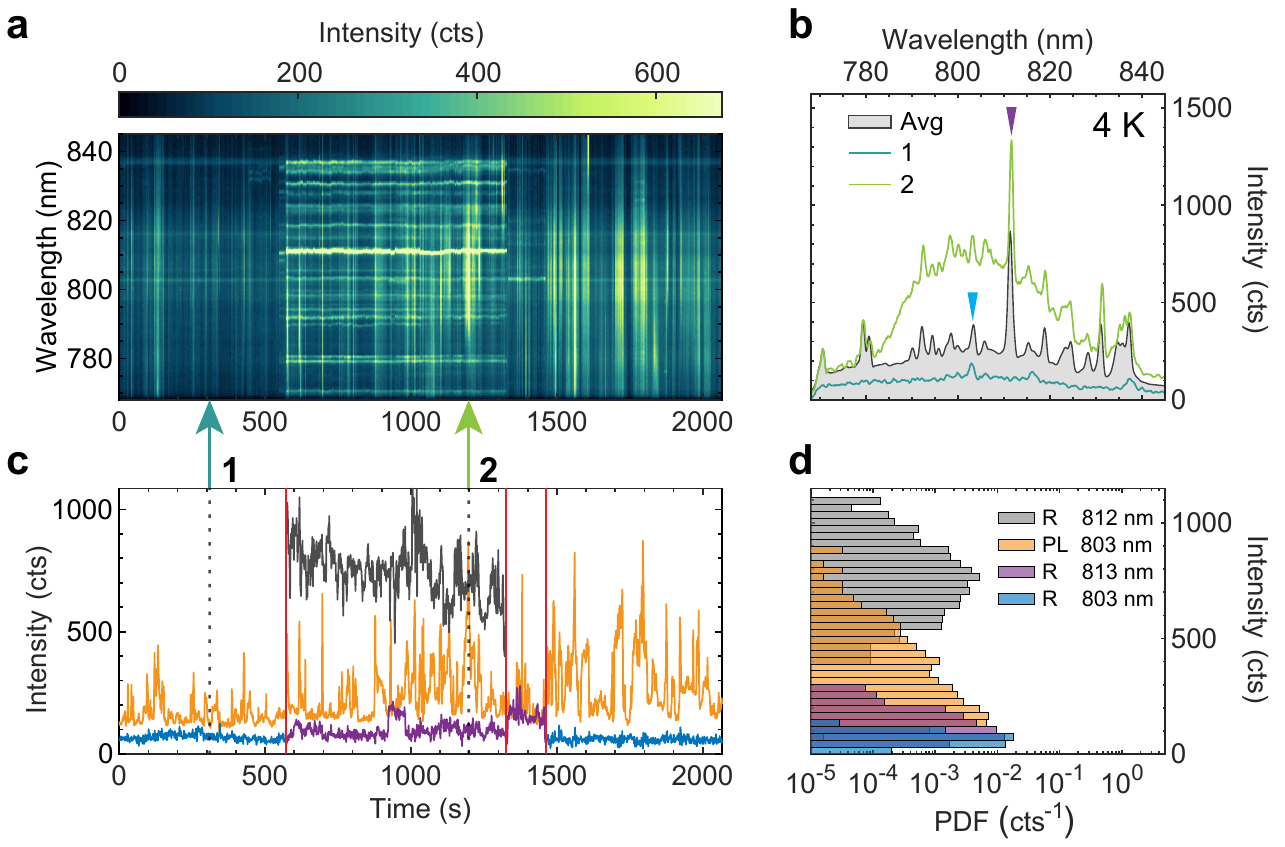}
    \caption{
    \textbf{Simultaneous PL and Raman measurement at cryogenic temperature.}
Different mechanisms for PL blinking and Raman fluctuations at 3.8 K (sample temperature) from sample No.~2 measured by the setup in Figure \ref{fig:Cryostat}. (\textbf{a}) Time series (color map) and (\textbf{b}) representative individual spectra of the emission from a single nanojunction under simultaneous 532 nm and 740 nm excitation with respective power densities $\sim9\times10^{3}$ W/cm$^{2}$ and $\sim1.3\times10^{2}$ W/cm$^{2}$, at 3.8~K sample temperature. A very long period of anomalous Raman signal (probably coming from one or few molecules) is observable, while the PL continues to blink independently of that.  (\textbf{c}) Intensity traces and (\textbf{d}) corresponding probability density functions (PDFs) for PL (orange) and Raman scattering (blue and purple for the peak at 803 nm, grey for the peak emerging temporarily at 811 nm). 
We do note a slightly increased blinking during, and persisting after, the anomalous Raman event, suggesting that the latter is caused by a metal restructuring that can also contribute to blinking. More research is needed to establish the relationship between anomalous Raman and blinking PL. 
    }
    \label{fig:PL_R_pico_4K}
\end{figure}
%-------------------------------------------------------------------------------------------------------------------------------------------------------------------------------------------------------

\clearpage

%\subsection*{Blinking activation by interband excitation}

\begin{figure}[htp!]
    \centering
    \includegraphics[width=0.8\columnwidth]{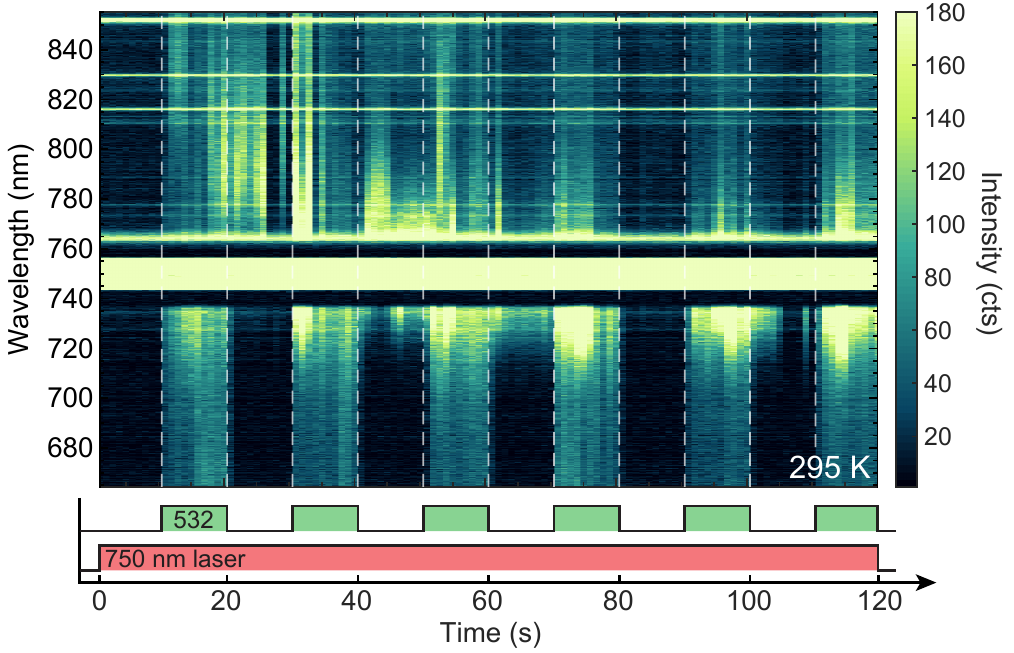}
    \caption{
    \textbf{Blinking activation by interband excitation.}
Two-color Raman+PL measurement with continuous excitation at 750 nm and intermittent excitation (periods of 10~s ON followed by 10~s OFF) at 532 nm (on sample No.2 in Supplementary Table~\ref{tab:Sample_list}). The experimental parameters are the same as in Fig.~3 in the main text.
At the start of the experiment under 750~nm excitation alone the background emission is virtually absent, only the Raman signal is visible. 
After the 532~nm illumination is switched ON for 10~s and OFF again, however, we observe a persistently increased background emission under 750~nm excitation alone, suggesting that the blinking emission centers have been activated by the green laser.
The result indicates that interband transitions and the generation of non-thermal photoexcited electrons play an important role in the activation of blinking.
This observation is in line with the temperature-dependent measurements shown in the main text and heat simulations presented below, as it supports a non-thermal activation channel of blinking emitters.
}
    \label{fig:Fig_ONOFF_analysis}
\end{figure}
%-------------------------------------------------------------------------------------------------------------------------------------------------------------------------------------------------------

\clearpage
%\subsection*{Temperature-dependent PL measurements}
%-------------------------------------------------------------------------------------------------------------------------------------------------------------------------------------------------------

\begin{figure}[htp!]
    \centering
    \includegraphics[width=0.9\columnwidth]{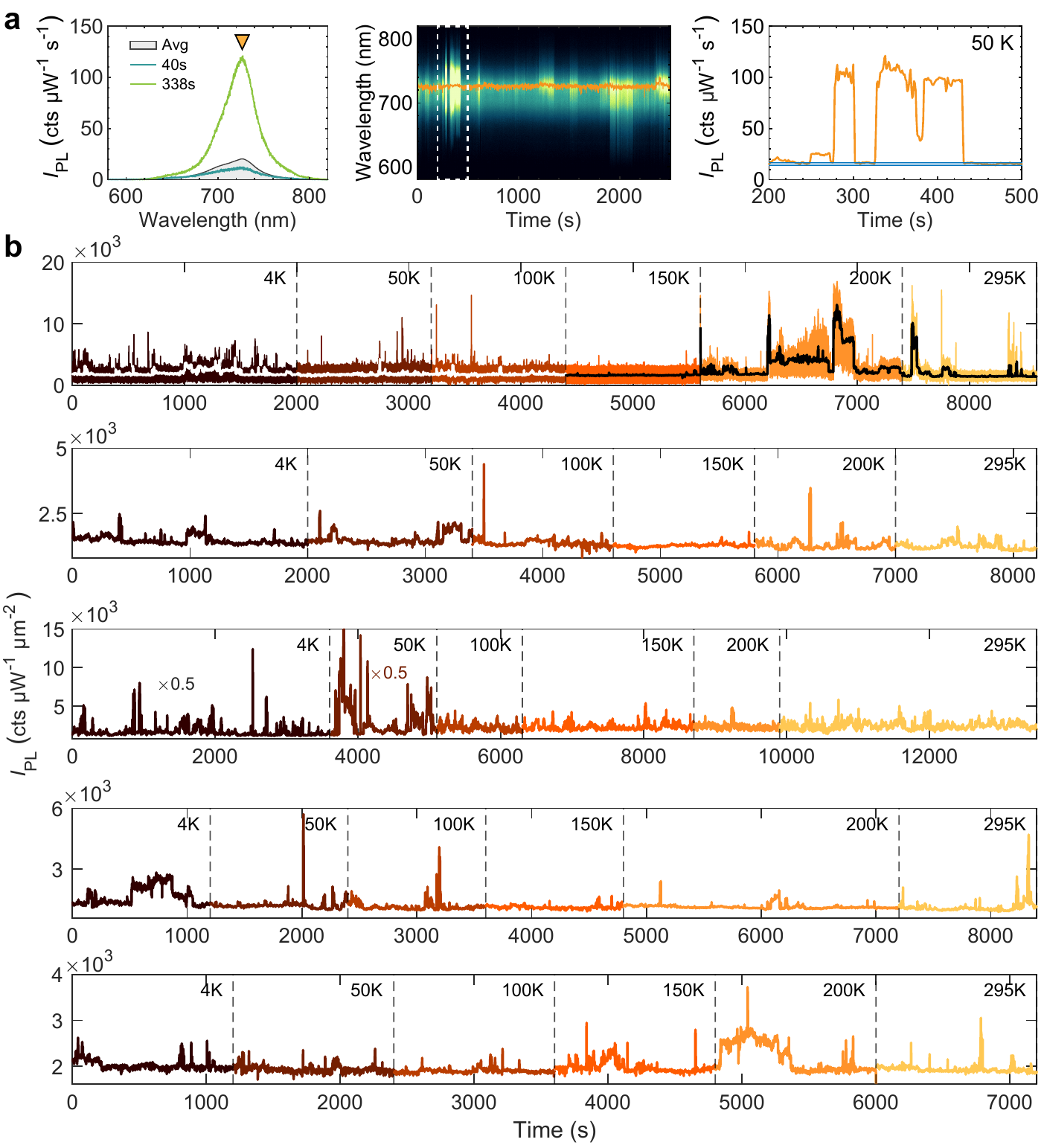}
    \caption{
    \small{\textbf{Temperature-dependent PL measurements.}
(\textbf{a}) Time series of PL, selected PL spectra and peak intensity trace from a single nanojunction (on sample No.~7 in Supplementary Table \ref{tab:Sample_list}) at 50~K by 532 nm excitation. Laser power, \SI{32}{\micro\watt}, integration time, 1~s.
% The right panel shows the long lasting blinking events, which tend to happen more frequently at cryogenic temperature than at room temperature. 
(\textbf{b}) Time series of the PL peak intensity from different individual nanojunctions (from No.~19 in Supplementary Table \ref{tab:Sample_list}) as a function of temperature measured by a single photon counting module. The first panel shows both 1~s and 1~ms binning time, while the traces in the remaining panel are binned to 1~s.
No clear correlation between the blinking behavior and sample temperature can be found, with occasional periods of increased blinking occurring at different temperatures, case by case. This strongly favour a mechanism that is not thermally activated.
    }}
    \label{fig:Temperature_dependence_Au}
\end{figure}
%-------------------------------------------------------------------------------------------------------------------------------------------------------------------------------------------------------

\begin{figure}[htp!]
    \centering
    \includegraphics[width=0.8\columnwidth]{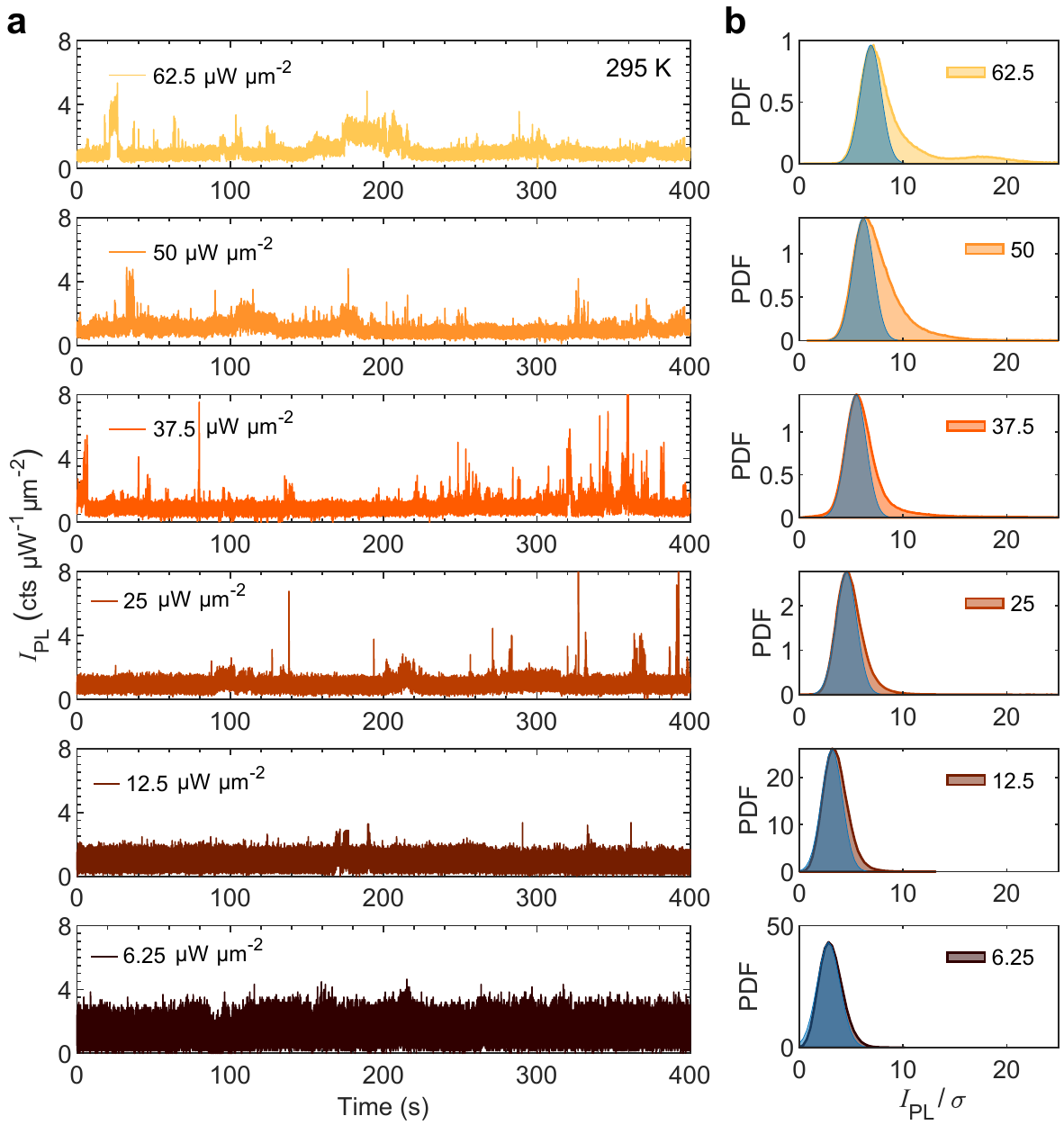}
    \caption{
    \textbf{Power-dependent PL measurements.}
Excitation power (at 532 nm) dependence of the PL intensity from a single nanojunctions (No.~7 in Supplementary Table \ref{tab:Sample_list}) at room temperature. 
(\textbf{a}) PL intensity traces at different laser powers collected by a single photon counting module with 1~ms binning time. 
(\textbf{b}) Corresponding probability density functions (PDFs) of the traces compared with PDF of the corresponding intrinsic measurement noise (dark blue shade). The measured counts $I_\mathrm{PL}$ is normalized by the standard deviation $\sigma$ to make the histograms comparable. Figure 5C in the main text is plotted by merging the six panels together.
    }
    \label{fig:power_dependence_Au}
\end{figure}
%-------------------------------------------------------------------------------------------------------------------------------------------------------------------------------------------------------

\clearpage
%\subsection*{{Multi-peak analysis of PL blinking from additional examples}}
%-------------------------------------------------------------------------------------------------------------------------------------------------------------------------------------------------------
\begin{figure}[htp!]
    \centering
    \includegraphics[width=0.8\columnwidth]{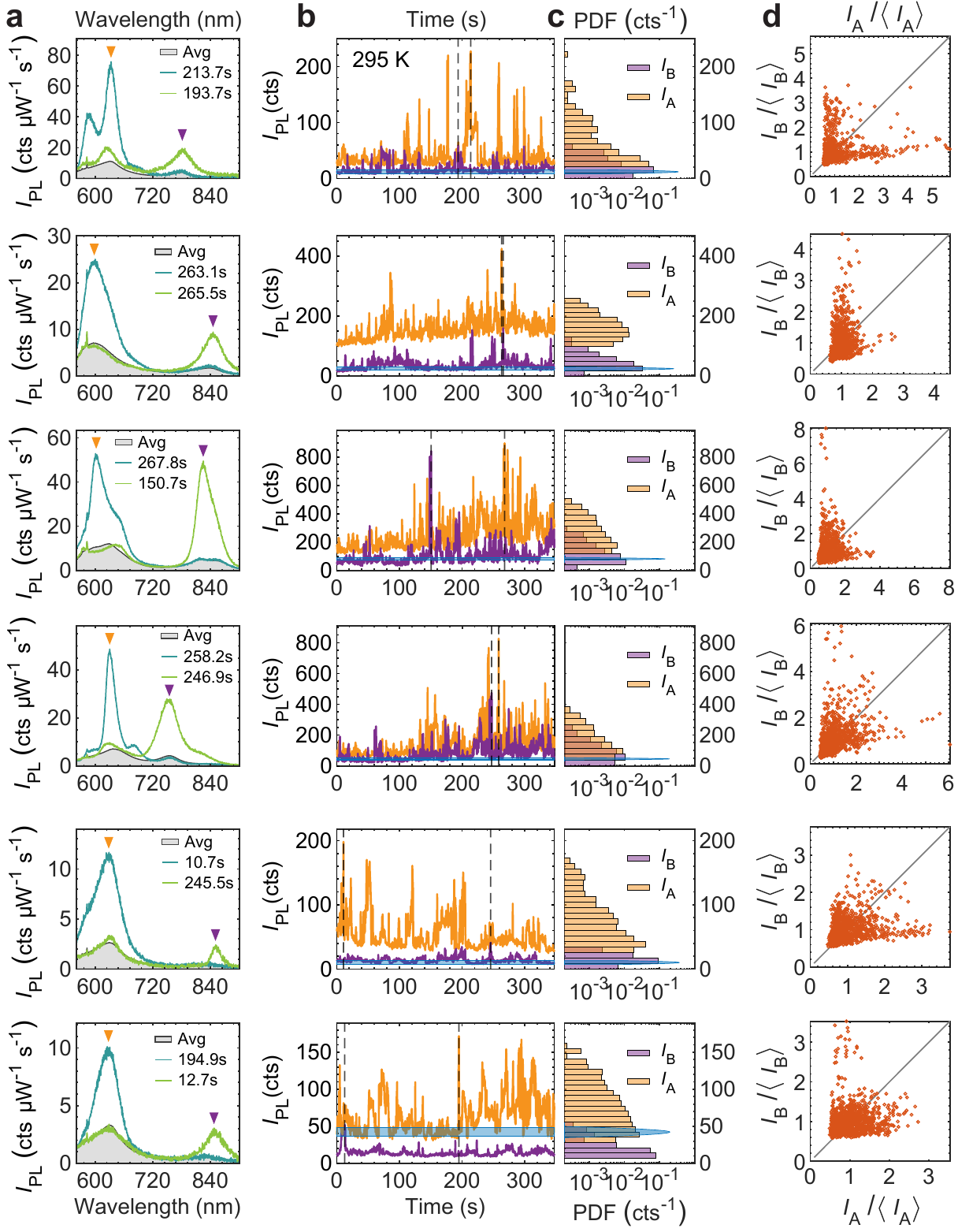}
    \caption{
    \textbf{Multi-peak analysis of PL blinking from additional examples.}
PL time series from 6 different individual nanojunctions under 532 nm excitation with 0.1s exposure time (from sample No.~3 in Supplementary Table \ref{tab:Sample_list} ). 
(\textbf{a}) Representative PL spectra of respective nanojunctions. 
(\textbf{b}) Intensity trace and (\textbf{c}) corresponding probability density functions
of the PL peak A ($<$700 nm, orange) and B ($>$700 nm, purple). 
Shaded blue areas represent the noise associated with each of these measurements. 
(\textbf{d}) Correlation map between the intensity fluctuations of peaks A and B. 
No clear correlation between the PL of the different plasmonic modes can be evidenced. 
    }

    \label{fig:PL_MP10}
\end{figure}
%-------------------------------------------------------------------------------------------------------------------------------------------------------------------------------------------------------

%4444444444444444444444444444444444444444444444444444444444444444444444444444444444444444444444444444444444444444444444444444444

\clearpage

\clearpage
%\subsection*{{Simulation of laser heating effect}}
%-------------------------------------------------------------------------------------------------------------------------------------------------------------------------------------------------------
\begin{figure}[htp!]
    \centering
    \includegraphics[width=0.7\columnwidth]{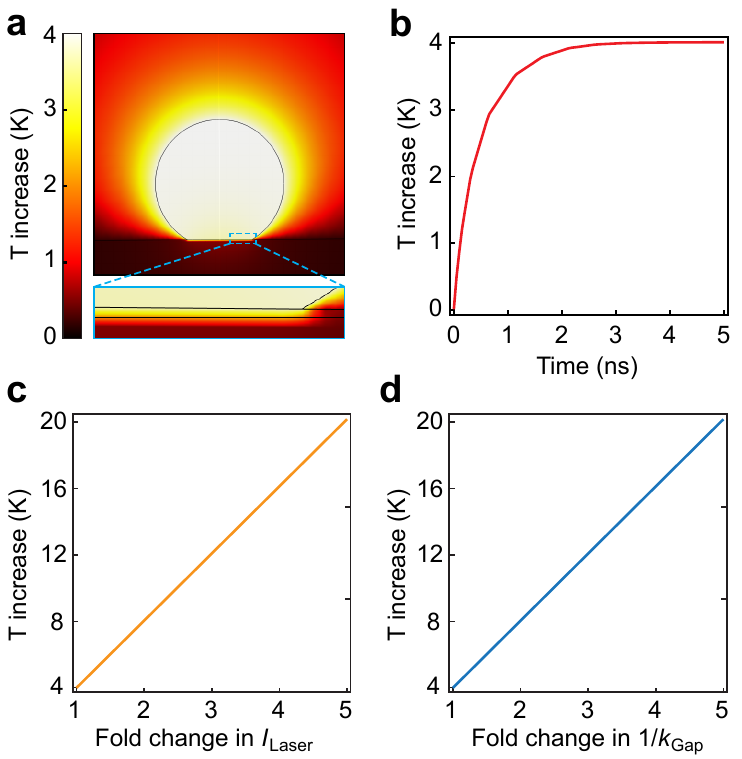}
    \caption{
%\begin{small} 
\textbf{Simulation of the temperature increase from a nanojunction by laser heating.}
(\textbf{a}) Temperature increase distribution with $\sim 3.5\times 10^4~$W/cm$^2$  light excitation at 532~nm. 
(\textbf{b}) Time-dependent temperature increase, showing saturation of the temperature increase after a few nanoseconds. 
(\textbf{c}) The heat absorption and temperature increase linearly by increasing laser power $I_\mathrm{Laser}$.
(\textbf{d}) Temperature increase as a function of thermal conductivity in the gap $k_\mathrm{Gap}$.
The small temperature increase results from the high thermal conductivity of the gold substrate. 
%\end{small}  
    }

    \label{fig:Heating_sim}
\end{figure}
%-------------------------------------------------------------------------------------------------------------------------------------------------------------------------------------------------------

\clearpage
%-------------------------------------------------------------------------------------------------------------------------------------------------------------------------------------------------------
\begin{figure}[htp!]
    \centering
    \includegraphics[width=0.4\columnwidth]{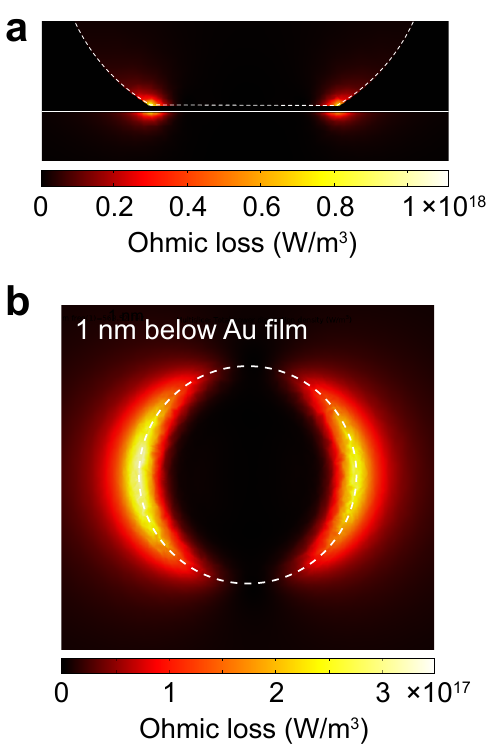}
    \caption{
    \textbf{Estimation of photon absorption rate by the nanojunction.}
Vertical cross-section (\textbf{a}) and in-plane (\textbf{b}, 1 nm below the Au film top surface) views of the simulated absorption distribution in a nanojunction illuminated with $3.5\times 10^4~$W/cm$^2$ power density of monochromatic light at 532~nm (largest power used in the experiment). All parameters are the same as in Supplementary Fig.~\ref{fig:Heating_sim} and are described in the preceding text.
The spatially integrated absorbed power from the gap region is $P_{\mathrm{Gap}} = 3.7\times 10^{-6}~$ W. We can translate this into the number of photons absorbed from the gap region within a duration $T$ through $N_\mathrm{Abs} = {P_{\mathrm{Gap}} T}/{\hbar \omega}$, where ${\hbar}$ is the reduced Planck constant and $\omega$ is the laser frequency at 532 nm. The resulting $N_\mathrm{Abs}$ is about 10 photons per picosecond, where 1 ps is the timescale for energy relaxation through electron-phonon interaction. The 532 nm laser power used in our experiment to observe blinking is generally 10 times smaller, which gives one absorbed photon per picosecond, as quoted in the text.
    }
    \label{fig:Heating_sim2}
\end{figure}
%-------------------------------------------------------------------------------------------------------------------------------------------------------------------------------------------------------

\clearpage
%\subsection*{Brief review of fluctuating emission from plasmonic hot-spots}
\subsection*{Supplementary Note 1: A brief review of fluctuating SERS continuum from plasmonic hot-spots in comparison with PL blinking}
%Blinking is defined as XXX, which has been well studied in various quantum emitter systems. 

In this section, we give a brief overview of previous observations of fluctuating emission from photo-excited plasmonic hot-spots, mainly in the context of surface-enhanced Raman scattering (SERS), and of the proposed mechanisms. 
We stress upfront that the vast majority of the literature on the topic is related to Raman signal fluctuations typically considered as a signature of single-molecule SERS (SM-SERS)
\cite{nie1997,kneipp1997,michaels1999,xu1999,bjerneld2000,michaels2000,otto2002,kudelski2007,stranahan2010,lombardi2011}. 
Overall, although a variety of models have been proposed to understand signal fluctuations in SM-SERS (and the origin of the gigantic signal enhancement), the underlying principles are still under debate due to the complex interactions between adsorbates and plasmonic hot-spots, and their evolution in the course of the experiment.
The samples used in previous research on SM-SERS were most frequently fabricated by the mixture of salt-aggregated Ag nanoparticles and a very low concentration of analyte molecules (rhodamine 6G, crystal violet, biomolecules, etc.) \cite{nie1997,kneipp1997,michaels1999,xu1999}. 
Without accurate control, it was found that the analyte randomly diffuses in and out of the hot-spot region and that the molecular orientation varies \cite{jiang2003,maruyama2004}, leading to fluctuations of the Raman scattering spectrum and intensity 
\cite{moyer2000,otto2001,weiss2001,futamata2002,emory2006,stranahan2010}.
Interestingly, the broad emission underlying the Raman peaks (called `SERS continuum'), which is absent from powders and pure molecule ensembles, was found to fluctuate together with the Raman signal in numerous previous reports
\cite{michaels1999,michaels2000,mihalcea2001,meixner2001,weiss2001,bosnick2002,andersen2004,
itoh2006a,bizzarri2007,weber2012}. 
While much research crystallized on elucidating the origin of the `SERS continuum', no single mechanism was yet demonstrated to account for all experimental data.
\\

On the contrary, in plasmonic nanojunctions with well-controlled geometry used in our work, metal luminescence blinking is not accompanied by measurable Raman signal fluctuations. 
Therefore, the above-mentioned phenomenology cannot be mapped onto our observations. 
%where metal luminescence blinking is not accompanied by measurable Raman signal fluctuations. 
On the other hand, owing to the similarity in the phenomenon, the `SERS continuum' fluctuations -- reported mostly from silver particles -- could be intuitively considered as sharing the same mechanisms as the PL blinking reported here in gold nanojunctions. 
Nevertheless, these two phenomena actually feature significant differences.  
% For instance, whereas luminescence and PL result from interband transition (usually by green light for Au, or ultraviolet light for Ag), SERS continuum is mainly attributed to intraband transitions or electronic Raman process, where the interband transition is not allowed or negligible.
To further clarify their relationship and differences, we review below the main hypotheses for the origin of the `SERS continuum' fluctuations, and assess them against our observation of PL blinking. 
\\

%Nevertheless, many SM-SERS studies, mostly conducted with silver particles, also reported background fluctuations, that could share similar mechanisms with PL blinking reported here in gold nanojunctions. 

%In contrast to the well controlled nanojunction geometry used in our work, samples used in previous research on SM-SERS were most frequently fabricated by the mixture of salt-aggregated Ag nanoparticles and a very low concentration of analyte molecules (rhodamine 6G, crystal violet, biomolecules, etc.) \cite{nie1997,kneipp1997,michaels1999,xu1999}. 
%Without accurate control, it was found that the analyte randomly diffuses in and out of the hot-spot region and that the molecular orientation varies \cite{jiang2003,maruyama2004}, leading to fluctuations of the Raman scattering spectrum and intensity 
%\cite{moyer2000,otto2001,weiss2001,futamata2002,emory2006,stranahan2010}. 

\paragraph{Main fluctuation mechanisms proposed to date} 
\label{review}
~\\

One possible origin for the intensity fluctuation of the `SERS continuum' is a variation of the local field enhancement, which can be caused either by a global effect involving  the  entire plasmonic structures
\cite{maruyama2004,emory2006,kudelski2006}
(e.g., change of the gap size) or a local effect involving adatom diffusion dynamics on the metal surface  \cite{benz2016a,shin2018}.
The latter was proposed to cause atomic-scale confinement of light and activate new Raman modes in the surrounding molecules due to the large field gradient (a phenomenon that we also report in Supplementary Fig.~\ref{fig:PL_R_pico_4K}).
As our simultaneous PL+Raman and PL+DF measurements demonstrate, such a mechanism cannot explain the pronounced intrinsic PL blinking that we report here.
In particular, the results of Fig. 2 in the main text suggests the independent generation of PL emission coupled to different plasmonic modes, which together with the PL+DF measurements (Fig. 4 in the main text) exclude any mechanism that would be linked with the change of the bulk plasmon frequency, as proposed in \cite{carnegie2020}.\\

A second possible fluctuation mechanism is based on dynamical charge transfer between adsorbate and metal, which is sometimes referred to as the chemical enhancement factor in SERS \cite{moskovits1985,otto2001,weiss2001,emory2006,kudelski2007,bizzarri2007,lombardi2011}.
In this model, spectral fluctuations are caused by chemical adsorption or desorption of the molecule, or by its thermally activated atomic scale movement.  
It was also proposed that such change of adsorbate-metal interaction can give rise to the fluctuation of the SERS continuum.
This model relates the SERS continuum to an electronic Raman process that is enhanced by the relaxation of momentum conservation when a chemisorbed molecule acts as a localised `defect' on the metal surface
\cite{gass1989,bosnick2002,jiang2003,monti2004,moore2005}. 
In light of our data, this mechanism could explain some blinking events in which a plasmonic mode becomes momentarily brighter while conserving its line shape and peak wavelength. 
However, by comparing the various types of samples we synthesized, we see that PL blinking also happens in the plasmonic nanojunctions without molecular chemisorption (e.g. dielectric and citrate spacers from samples No. 17 and 19 in Supplementary Table~\ref{tab:Sample_list}).
As a result, we conclude that if an electronic Raman process is at play during some blinking periods, then it can also be induced by intrinsic defects in the metal surface layers, and not only by adsorbed moieties. \\

Recently, the variation of bulk plasma frequency induced by local defects on the metal interface was proposed as a mechanism to explain the fluctuating `SERS continuum' under stable Raman signal in plasmonic hot-spots \cite{carnegie2020}. 
Although this phenomenon seems at first similar to our results from the two-color PL+Raman measurement (Fig.~3 in the main text), the `SERS continuum' fluctuations in \cite{carnegie2020} are predicted to be correlated with a pronounced shift of the entire plasmonic resonance spectrum. It is at odds with our observations of stable DF scattering spectrum during blinking.  
On the other hand, the authors of \cite{carnegie2020} explain the `SERS continuum' as electronic Raman scattering rather than electron-hole recombination process, which can apply to measurements under near-infrared excitation (Fig. S10c and S10d) but does not represent the dominant interband transition processes at play under green excitation. 
\\

Another possible source of spectral fluctuation is the contamination of amorphous carbon around the plasmonic hot-spots, which generally comes from the damage of carbon-based molecules under laser irradiation or heating from analyte, impurities from the solution, or even from the air (e.g., CO)
\cite{tsang1980,kudelski2000,moyer2000,otto2002,kudelski2004,monti2004,lucotti2006,kudelski2006,
kudelski2007,chaigneau2010}. 
Amorphous carbon shows quite broad Raman bands with the strongest two peaks around 1300 cm$^{-1}$ and 1600 cm$^{-1}$ and their overtone modes around 3000 cm$^{-1}$, which are also recognised as a type of SERS continuum \cite{mrozek1990,moyer2000,robertson2002}. 
In our experiment, the stable Raman signal with low background under near-infrared light excitation (Fig.~3 in the main text) suggests that carbon contamination does not have any significant contribution to the PL blinking. 
This conclusion is supported by the high sensitivity of blinking behavior on spacer material, even though all samples are expected to be equally contaminated by amorphous carbon since they are prepared and studied in similar conditions. 
%in particular for the measurements in the cryostat, and also because our nanojunctions are capped by an alumina layer.
%On the other hand, the carbon contamination is generally ubiquitous and unavoidable.
If PL blinking were related to carbon contamination, we believe that the prominence and characteristics of PL blinking would be the same for all samples, which is clearly disproved by our PL measurements (Supplementary Fig.~\ref{fig:comparison} and \ref{fig:more_type}).
\\

To place our work into context, it should be emphasized that most SERS blinking observations were reported from Ag systems, which show pronounced chemical reactivity
\cite{peyser2001,mihalcea2001,peyser2002,andersen2004,monti2004,jacobson2005,
jacobson2006,wu2008,borys2011}, in contrast to gold which is a rather inert substrate at ambient conditions.
Indeed, luminescent Ag adatoms can be photochemically generated from a silver oxide system under laser irradiation \cite{peyser2001,peyser2002,andersen2004,monti2004}.
Combining this effect with oxidation from Ag to Ag oxide in air, a reversible photochemical reaction loop can be realised, resulting in luminescence blinking even from a bare Ag system without any Raman probe \cite{andersen2004,jacobson2005,jacobson2006,wu2008}.
In parallel, Ag can also exhibit a strong interaction with CO from the environment; 
the Raman signal of carbon contaminants can even be found from fresh Ag films deposited under high vacuum conditions \cite{mrozek1990,kudelski2004}. 
All these effects make the blinking phenomenon in Ag systems easier to observe but more complicated to analyze compared to Au systems, which is why we focused our main study on gold. 
There is rare literature reporting luminescence fluctuations from Au systems, but existing reports either lack control over the structures and corresponding local fields \cite{geddes2003a} or independent characterisation of the plasmonic response and possible changes in the local fields \cite{li2017a}.\\

Clarifying the underlying mechanisms relies on developing a stable plasmonic platform with much better control of morphology compared to previous systems, as demonstrated in our nanoparticle-on-mirror systems. In contrast with previous reports where the continuum emission fluctuated together with the Raman signal
\cite{michaels1999,michaels2000,mihalcea2001,meixner2001,weiss2001,bosnick2002,andersen2004,itoh2006a,bizzarri2007,weber2012} 
(or without discrimination between them), 
our observation of PL blinking without Raman blinking clearly evidences that other changes are occurring within the metal instead of inside the gap. 
This excludes mechanisms involving molecule diffusion or local field variation (such as `picocavities'). 
%Carbon contamination can also be excluded as a cause for PL blinking, in particular for the measurements in the cryostat, and also because our nanojunctions are capped by an alumina layer. 
More generally, unlike in Ag systems, it is improbable that any photochemical reactions take place between gold and the immediate environment \cite{boyen2002,tsai2003,ono2008}.
Luminescence blinking is also unlikely to arise from the `off-resonant' molecules (i.e., do not absorb nor luminescence at the relevant wavelengths) used in our plasmonic nanojunctions. 

%On the contrary, our model based on Purcell enhanced PL by localised emitters not only fit the observation of Raman and DF, but also compatiable with the ***

%The fact that we see the blinking PL coupling to specific modes tells us about the location of the blinking emitter, which has to be inside the gap region. This restricts possible models, since it is not possible to have many layers of adatoms forming a cluster in the gap without affecting the Raman signal or DF. 

%{carnegie2020} 

%5555555555555555555555555555555555555555555555555555555555555555555555555555555555555555555555555555555555555555555555555555555

\clearpage
\subsection*{Supplementary Note 2: Comparison of PL blinking from different types of nanojunctions}
\label{sample comparison}

To test the generality of the blinking phenomenon, and to investigate the impact of nanojunction composition on blinking, we fabricated and characterised many different types of nanojunctions (see Supplementary Table~\ref{tab:Sample_list}), and some measurement examples are presented in Supplementary Fig.~\ref{fig:comparison} and Supplementary Fig.~\ref{fig:more_type}. 
As explained above, we systematically changed the substrate type, the spacer layer, and the nanoparticle material and shape, while maintaining similar plasmonic resonance frequencies and mode volumes.
A general conclusion can be drawn from these measurements: the occurrence frequency, duration and strength of the PL fluctuations are governed mostly by the metal surface chemistry; on the contrary, PL blinking seems to weakly depend on the degree of crystallinity of the substrate, or on the composition and shape of the nanoparticle forming the junction. 
This highlights the key role of photo-induced restructuring of the metal surface layer.

Fluctuations are found to be more pronounced in the nanojunctions with molecules linked by thiol groups to the Au film surface (Supplementary Fig.~\ref{fig:comparison}a,~\ref{fig:more_type}a, \ref{fig:more_type}b, \ref{fig:more_type}d and \ref{fig:more_type}f), while the structures with an oxide layer separating the Au film from the molecules tend to yield more stable emission, with only short bursts lasting less than one second (Supplementary Fig.~\ref{fig:comparison}b and \ref{fig:comparison}c).
This suggests the significantly larger contribution from Au film to the blinking emission than from the nanoparticles.

When using a single layer MoS$_2$ as a robust crystalline spacer, which strongly binds to both the Au nanoparticle and film by sulfur atoms (Supplementary Fig.~\ref{fig:comparison}e), PL fluctuations are not as pronounced as with Au-molecule-Au configurations.
This highlights the role of molecular species in destabilising the metal surface layer and promoting light-induced cluster formation. \\

Silver nanojunctions blink much more strongly than Au nanojunctions, even though the Ag film is covered by a compact oxide layer, and only native ligand molecules may remain (Supplementary Fig.~\ref{fig:comparison}f). 
This may be because the chemical reactivity of Ag and the field enhancement are  higher than those of Au, which probably boost the formation of atomic clusters responsible for blinking emission.
Remarkably, the emission enhancement reaches almost 6 orders of magnitude compared to the substrate emission, which corresponds to approximately 8 to 9 orders of magnitude enhancement factor when accounting for the size of the nanojunction. 
The resulting PL quantum yield seems on par with that of semiconducting emitters (but we did not perform a quantitative quantum yield measurement).\\

In order to clarify the relationship between PL blinking and the existence of a nano-gap yielding extreme field confinement, we study the PL of an individual 150-nm-diameter Au nanoparticle on a 100-nm-thick SiO$_2$ layer over a Si substrate (Supplementary Fig.~\ref{fig:NP150}).
As shown in Supplementary Fig.~\ref{fig:NP150}d and \ref{fig:NP150}e, both simulated and measured scattering spectra show a broad plasmonic resonance covering 600-700 nm, which can be identified as a dipole mode based on the charge distribution (inset in Supplementary Fig.~\ref{fig:NP150}d). 
Under s-polarised excitation, the dipole mode is equivalent to the transverse mode of the nanojunction. The dipole mode shows a red-shift under p-polarised excitation (Supplementary Fig.~\ref{fig:NP150}d and \ref{fig:NP150}e) because more charge is localised in the bottom facet causing stronger coupling with the substrate.
An additional peak near the excitation wavelength (beneath the Si Raman peaks) can be found in the PL spectrum (Supplementary Fig.~\ref{fig:NP150}e), which originates from the edge of interband transitions in gold \cite{yorulmaz2012a,frohlich2016,cai2018}. 

An example of PL time-series measurement is shown in Supplementary Fig.~\ref{fig:NP150}e and \ref{fig:NP150}f.
While the PL spectra of the individual nanoparticle are generally stable compared to the nanojunctions, we do observe rare and weak blinking events (Supplementary Fig.~\ref{fig:NP150}g).
Note that citrate molecules are loosely covered on the nanoparticle, which is in line with our proposed mechanism that the presence of molecules at the surface plays an important role in facilitating adatom migration and the generation of atomic defects.
This very low blinking occurrence may also explain why PL blinking effect was not reported from most-studied individual metal nanoparticles (e.g. nanorods) \cite{fang2012,cai2018}.
From this comparison, we conclude that the much higher field enhancement provided by the nanoparticle-on-mirror (at least one order of magnitude larger than at the surface of isolated particles) is essential in triggering the blinking mechanism.

%\section*{References}
%\nocite{*}

\end{document}